%% file: TSP_template.tex
\documentclass[journal,article,submit,moreauthors,pdftex]{Definitions/tsp} 
\input{Definitions/package}
\include{Definitions/unicode}

\continuouspages{}
\firstpage{1} 
\pubvolume{0}
\issuenum{0}
\elocationid{0}
\articlenumber{0}
\pubyear{2026}
\copyrightyear{2026}
\datereceived{Day Month Year}
\dateaccepted{Day Month Year}
\dateonlinefirst{}
\datepublished{Day Month Year}
\usepackage[OT1,OT2,T2A,T2B,T2C,T3,T5,T1]{fontenc}
\usepackage[russian,english]{babel}
\usepackage{geometry}
\usepackage{pgfplots}
\usepackage[T1]{fontenc}
\usepackage[utf8]{inputenc}
\usepackage{booktabs, longtable, makecell, multirow, xltabular}
\usepackage{verbatim}

\Title{Ember: A Serverless Peer-to-Peer End-to-End Encrypted Messaging System over an IPv6 Mesh Network}
\Author{Hamish Alsop\textsuperscript{1,*}, Leandros Maglaras\textsuperscript{1,2, *} and Naghmeh Moradpoor\textsuperscript{1}}
\AuthorNames{Hamish Alsop, Leandros Maglaras, Naghmeh Moradpoor}

\address{%
\textsuperscript{1} School of Computing, Edinburgh Napier University, Edinburgh, UK

\textsuperscript{2} School of Computing, De Montfort University, Leicester, UK
}%
\corres{Corresponding Authors: Hamish Alsop. Email: 40342915@napier.live.ac.uk, Leandros Maglaras, Email: leandros.maglaras2@dmu.ac.uk}

\abstract{A substantial body of research has focused on formalising what constitutes a ``secure'' messaging system, recognising that end-to-end encryption alone is insufficient to capture the full range of security, privacy, and usability properties that are expected by modern users. Several solutions have been proposed recently, including their own drawbacks, making the need for a direct secure messaging system a necessity. This paper presents \emph{Ember}, a serverless peer-to-peer messaging system providing end-to-end encrypted communication over a decentralised IPv6 mesh network. Ember operates without central servers, enforces data minimisation through ciphertext-only local storage and time-based message expiration, and prioritises architectural clarity, explicit trust boundaries, and practical deployability on Android. The paper describes the system architecture, cryptographic design, network model, and security properties.  Ember includes a ciphertext-only persistence model using an encrypted local database, ensuring that message plaintext is never written to disk. Through the integration of a time-to-live (TTL) mechanism for automatic message expiration, Ember enforces data minimisation on mobile devices. Ember is a layered and analysable system architecture with explicit trust boundaries separating user interface logic, cryptographic operations, storage, and networking components. The paper presents dynamic testing results demonstrating that no plaintext information can be recoverable from captured network traffic, and finally discusses limitations and future work. }


\begin{document}

\section{Introduction} \label{sect:s1}
End-to-end encrypted (E2EE) messaging has become foundational to modern digital communication, but its widespread deployment has attracted increasing regulatory attention. Proposed legislative initiatives---most notably the European Union's ``Chat Control'' framework---seek to introduce mechanisms for lawful access to private communications, raising serious concerns within the security research community about the long-term viability of strong E2EE guarantees \cite{akalin2026chat}.

Such proposals typically rely on client-side content scanning or mandatory interception capabilities that, if deployed, would negate the end-to-end security model by introducing surveillance at the endpoint. From a security engineering perspective, these approaches reintroduce systemic vulnerabilities, create new attack surfaces, and erode the trust assumptions on which secure messaging is built \cite{abelson2024bugs}.

In parallel, most widely deployed messaging platforms rely on centralised architectures where message routing, metadata storage, and key management are mediated by service providers. Even where encryption protects message contents, users must place substantial trust in these providers and in the legal environments in which they operate \cite{alsop2025innovating}. Centralisation introduces single points of failure---both technical and institutional---and concentrates the risk of large-scale surveillance, data retention, or compelled access.

This work is motivated by two questions. First, can a secure messaging system be constructed in a manner that meaningfully resists regulatory and infrastructural pressures on end-to-end encryption? Second, can reliance on centralised service providers be reduced or eliminated entirely through decentralised, serverless communication? By removing central servers and minimising persistent data on end-user devices, this project explores an alternative design space that prioritises user autonomy, data minimisation, and architectural resilience.

The objective is not to deploy a mass-market platform, but to rigorously examine whether decentralised and serverless secure messaging remains technically viable under tightening legislative and infrastructural constraints.

\subsection{Problem Statement}

The problem addressed in this paper is the design and evaluation of a secure messaging system that operates without centralised infrastructure, minimises retained user data, and maintains end-to-end confidentiality on mobile platforms. In the context of emerging regulatory pressures on encrypted communication, there is a need to investigate architectures that avoid the traditional client--server model entirely and thereby reduce the applicability of legislation that assumes the presence of service providers or central intermediaries.

Specifically, this work examines whether peer-to-peer secure messaging can be achieved in a practical and analysable manner using direct device-to-device communication over a decentralised IPv6 mesh network. The absence of central servers eliminates conventional message queues, metadata aggregation points, and provider-controlled storage, fundamentally altering the threat model and trust assumptions compared to mainstream messaging applications.

Additionally, mobile platforms impose unique constraints, including background execution limits, restricted networking capabilities, and persistent storage risks. Any viable solution must therefore operate within these constraints while still enforcing strong security properties, such as encrypted storage, ephemeral message retention, and explicit key management.

The Ember system is designed to explore this problem space by implementing a serverless, peer-to-peer messaging architecture that provides end-to-end encryption, encrypted local persistence, and automatic message expiration. The system intentionally limits supported features, such as multimedia transfer and group messaging, in order to prioritise security clarity, correctness, and minimisation of retained data.

\subsection{Research Objectives}

The primary objective of this research is to assess whether the implemented Ember system satisfies its intended security and architectural goals under realistic operating conditions. Rather than proposing a novel cryptographic primitive, this work focuses on system-level design, integration, and validation of existing cryptographic mechanisms within a decentralised mobile environment.

The research objectives are as follows. First, to evaluate the confidentiality and integrity of message data transmitted over a peer-to-peer IPv6 mesh network, ensuring that message contents remain protected against network-level adversaries. Second, to validate that message data is securely stored at rest using encryption and that plaintext is not persistently retained on the device beyond volatile memory usage. Third, to assess the effectiveness of the system's time-to-live (TTL) mechanism in enforcing automatic message deletion and supporting data minimisation goals.

Further objectives include examining the practicality of identity verification via cryptographic key fingerprints in a serverless context, and evaluating the operational reliability of peer-to-peer communication over a decentralised IPv6 mesh network. In particular, this work seeks to determine whether such a networking model is a viable alternative to traditional server-mediated message delivery or whether it introduces prohibitive complexity or reliability challenges.

Importantly, this research explicitly does not target the implementation or validation of ratcheting-based post-compromise security (PCS) or forward secrecy (FS) guarantees, such as those provided by Double Ratchet protocols. While such mechanisms are well-established in centralised secure messaging systems, their correct integration in a serverless, peer-to-peer mobile environment introduces significant complexity and was not achieved within the scope of this project. As such, ratcheting-based guarantees are treated as future work rather than completed research objectives.

\subsection{Contributions}

This paper makes the following concrete and implemented contributions:

\begin{itemize}
    \item The design and implementation of a functioning serverless peer-to-peer messaging system that provides end-to-end encrypted communication over a decentralised IPv6 mesh network.
    \item A ciphertext-only persistence model using an encrypted local database, ensuring that message plaintext is never written to disk.
    \item The integration of a time-to-live (TTL) mechanism for automatic message expiration, enforcing data minimisation on mobile devices.
    \item The implementation of a key rotation protocol with mutual confirmation and HKDF-based key derivation, allowing controlled key evolution without loss of historical message decrypt-ability.
    \item A layered and analysable system architecture with explicit trust boundaries separating user interface logic, cryptographic operations, storage, and networking components.
\end{itemize}

These contributions collectively demonstrate the feasibility of decentralised, serverless secure messaging on mobile platforms while maintaining clear and defensible security guarantees.

The remainder of this paper proceeds as follows. Section~2 reviews related work. Section~3 defines the threat model. Sections~4--9 describe the system architecture, network design, cryptographic mechanisms, key lifecycle, data persistence, and message processing pipelines. Section~10 analyses privacy and metadata considerations. Section~11 presents security testing and verification results. Section~12 evaluates the system and discusses design trade-offs. Section~13 addresses limitations and future work, and Section~14 concludes.

\section{Related Work and Background}

\subsection{Secure Messaging Threat Models and Security Objectives}

A substantial body of research has sought to formalise what actually constitutes a ``secure'' messaging system, recognising that end-to-end encryption alone is insufficient to capture the full range of security, privacy, and usability properties that are expected by modern users. One of the most influential contributions in this area is the systematization of knowledge (SoK) on secure messaging by Unger et al.~\cite{sok_secure_messaging_2015}, which establishes a structured taxonomy of security goals, threat models, and design trade-offs across contemporary messaging applications.

The SoK identifies a set of cryptographic and system-level objectives that are now widely treated as baseline requirements. These include confidentiality and integrity under an active network adversary, authentication of communicating parties, forward secrecy (FS) to limit retrospective compromise, post-compromise security (PCS) to enable recovery after key exposure, and varying notions of deniability to be able to prevent cryptographic transcripts from serving as verifiable proof of authorship. The authors emphasise that no deployed system simultaneously achieves all desirable properties. Secure messaging design is therefore framed as an exercise in prioritisation, where stronger guarantees in one dimension often introduce complexity, usability costs, or new failure modes elsewhere.

Beyond payload confidentiality, the SoK explicitly elevates metadata to a first-class privacy concern. Even when message contents are end-to-end encrypted, information such as communication patterns, timing, peer relationships, and online presence could still remain observable to service providers or network adversaries. The work distinguishes between content security and metadata privacy, noting that most widely deployed systems strongly prioritise the former while offering limited protection for the latter. This distinction is particularly relevant for decentralised and peer-to-peer systems, where reducing central trust can introduce new forms of observability at the network or transport layers.

More recent work has reinforced and extended these observations. A 2025 systematic literature review of secure instant messaging applications~\cite{slr_secure_instant_messaging_2024} surveys a broad range of academic and applied systems, with particular emphasis on forensic artefacts, data persistence, and post-compromise recoverability. Although it is more motivated by forensic analysis, the review provides a comprehensive comparative overview of modern secure messaging architectures and highlights recurring gaps between theoretical security claims and real-world behaviour. Notably, it shows that many systems advertising ephemerality or ``disappearing messages'' continue to leave recoverable artefacts at the application, operating system, or storage layers, complicating strong claims around data minimisation and deletion.

These works collectively establish a practical analytical framework for evaluating secure messaging systems: content confidentiality must be weighed alongside compromise resilience, metadata exposure, architectural trust assumptions, and platform-level constraints. This framework directly informs Ember's design and evaluation. Rather than pursuing comprehensive metadata anonymity or maximal cryptographic guarantees, Ember prioritises high-assurance content confidentiality, reduced reliance on centralised infrastructure, and aggressive application-level data minimisation. The choice of a serverless, peer-to-peer architecture reflects an explicit trade-off: decentralisation reduces dependence on trusted servers but introduces challenges related to routing visibility, availability, and peer discovery. Situating Ember within these established threat models allows its design decisions to be evaluated transparently against prior work rather than in isolation.

\subsection{End-to-End Encryption Protocols and Ratcheting-Based Security}

Modern secure messaging systems predominantly derive their cryptographic strength from the Signal protocol family, which combines an asynchronous authenticated key exchange with a continuous key-evolution mechanism known as the Double Ratchet. At a high level, the Signal design decomposes into two principal components: the Extended Triple Diffie--Hellman (X3DH) handshake, which establishes an initial shared secret between parties that may not be simultaneously online, and the Double Ratchet protocol, which incrementally evolves cryptographic state to provide forward secrecy (FS) and post-compromise security (PCS) during message exchange.

Due to its central role in providing strong compromise resilience, the Double Ratchet has been the subject of extensive formal analysis. Collins et al.~\cite{tight_security_double_ratchet_2024} provide a tight security analysis of the protocol in the multi-session setting, addressing limitations in earlier proofs where security bounds degraded with the number of epochs or concurrent sessions. By decomposing the Double Ratchet into modular primitives---notably continuous key agreement (CKA) and forward-secure authenticated encryption (FS-AEAD)---their work clarifies which cryptographic assumptions underpin Signal's recovery guarantees following state exposure. Importantly, the analysis also shows that achieving tight reductions under standard Diffie--Hellman assumptions remains challenging, particularly at the scale and concurrency levels seen in real-world deployments.

Complementary work has examined ratcheting protocols using logic-based formal methods. Li et al.~\cite{formal_analysis_signal_loet_2024} apply Logic of Events Theory (LoET) to model both the X3DH handshake and the Double Ratchet phases of Signal, demonstrating that security guarantees are tightly coupled across protocol stages. Xiao et al.~\cite{formal_analysis_ratchet_mdpi_2025} generalise this approach to reason about ratcheting protocols as a class, introducing dedicated abstractions for key evolution, causality, and authentication. Collectively, these works highlight that ratcheting protocols are difficult to verify compositionally: guarantees provided by later stages depend critically on assumptions made during initial key establishment, and cannot be inferred just by analysing individual components in isolation.

The security of ratcheting-based designs is therefore inseparable from the properties of the underlying authenticated key exchange. X3DH remains the dominant design for asynchronous secure messaging, but has received comparatively less focused analysis. Hashimoto et al.~\cite{x3dh_efficient_construction_2022} recast X3DH as a Signal-conforming authenticated key exchange and present a generic construction based on standard primitives, including post-quantum key encapsulation mechanisms. Their analysis exposes fundamental trade-offs between efficiency, forward secrecy, and deniability, showing that design choices such as reuse of handshake material---made for performance and deployability---can measurably weaken forward secrecy guarantees.

Deniability introduces a further axis of complexity. Unger and Goldberg~\cite{strongly_deniable_ake_2016} show that achieving strong offline and online deniability in authenticated key exchanges often conflicts with resistance to key-compromise impersonation attacks and practical deployability constraints. More broadly, this body of work demonstrates that forward secrecy, post-compromise security, deniability, efficiency, and implementation complexity cannot all be maximised simultaneously; real-world systems inevitably need to make explicit trade-offs between these properties.

The practical upshot is clear: while ratcheting-based designs provide exceptionally strong guarantees under well-defined threat models, they rely on intricate state machines, strict key-management assumptions, and carefully coordinated interactions between protocol stages. Incorporating a Double Ratchet into a new system---particularly one with a decentralised or serverless architecture---is non-trivial and demands rigorous re-evaluation of threat models, trust assumptions, and failure modes.

Ember accordingly adopts a deliberately constrained cryptographic design. Rather than implementing a full ratcheting protocol, Ember employs per-contact symmetric keys combined with authenticated encryption to provide strong message confidentiality and integrity. This approach protects message contents against network adversaries and untrusted intermediaries, but does not provide continuous forward secrecy or post-compromise recovery across compromise windows. The trade-off is intentional: by avoiding complex ratcheting state and asynchronous handshake dependencies, Ember reduces implementation risk and aligns more closely with its serverless, peer-to-peer architecture. The extensive formal literature on Signal-style ratcheting therefore serves both as a benchmark for future extensions and as a justification for treating advanced ratcheting mechanisms as future work rather than baseline assumptions.

\subsection{Group Messaging and Multi-Device State}

While two-party secure messaging protocols such as Signal can provide strong guarantees through continuous ratcheting, extending these guarantees to group messaging and multi-device scenarios introduces substantially greater complexity. Group conversations are inherently dynamic: participants may join or leave at arbitrary times, users may operate multiple devices concurrently, and cryptographic state must remain consistent across all active members. Achieving forward secrecy (FS) and post-compromise security (PCS) under these conditions requires scalable mechanisms for group membership management, key evolution, and authentication.

The Messaging Layer Security (MLS) protocol represents the current consensus solution to these challenges. Standardised by the IETF as RFC~9420~\cite{rfc9420_mls}, MLS specifies an asynchronous group key establishment protocol designed to support groups ranging from a small number of participants to thousands, while maintaining FS and PCS across membership changes. Its design replaces ad hoc sender-key distribution with a structured tree-based group state, allowing cryptographic material to be updated efficiently as group membership evolves.

Importantly, MLS is not a self-contained messaging system. The MLS Architecture~\cite{rfc9420_mls} makes explicit that secure deployment depends on additional system components, including an authentication service for binding identities to credentials and a delivery service for reliable message dissemination. Correct operation further relies on application-level logic to handle group consistency, access control, and recovery from partial failures. Subsequent work, such as the formally verified TreeSync protocol by Wallez et al.~\cite{authenticated_group_management_mls_2023}, demonstrates that many historical weaknesses in group messaging systems arise not from cryptographic primitives themselves, but from underspecified group state management semantics. While this work strengthens confidence in MLS as a standard, it also illustrates the significant implementation and verification complexity involved.

These assumptions complicate the integration of MLS into fully decentralised or serverless environments. MLS presumes globally consistent group state, dependable message delivery, and a mechanism for authenticating long-term identities—properties that are straightforward to realise in centralised or federated systems, but considerably more difficult to achieve without reintroducing coordinating infrastructure. In peer-to-peer settings, challenges such as state synchronisation, denial-of-service resistance, and identity binding become first-order concerns rather than implementation details.

MLS accordingly serves as a benchmark for any future group extension rather than an immediate implementation target. Ember currently focuses on one-to-one communication over a serverless peer-to-peer overlay, prioritising minimal trusted infrastructure, reduced retained state, and analysable security properties. Supporting MLS-style group messaging or multi-device consistency would require substantial re-evaluation of Ember's trust assumptions and architectural goals. As such, group messaging is treated as explicit future work, with MLS providing a well-defined benchmark against which any future extensions can be evaluated.

\subsection{Serverless and Peer-to-Peer Secure Messaging Systems}

A smaller but influential body of work explores secure messaging systems that avoid centralised servers entirely, instead relying on direct peer-to-peer communication or minimal supporting infrastructure. These systems are typically motivated by strong censorship-resistance and metadata minimisation goals, and therefore provide the closest architectural comparison points to Ember.

Ricochet is one of the earliest practical implementations of serverless secure messaging. By leveraging Tor onion services as stable, pseudonymous endpoints, Ricochet enables direct peer-to-peer communication without revealing network identifiers or relying on central message relays. Kirsh's analysis of Ricochet's message-layer security~\cite{ricochet_message_layer_encryption} highlights both the strengths and limitations of this approach. While Tor provides transport-level anonymity and protects against local network observers, early Ricochet designs relied primarily on channel-layer encryption, leaving message content dependent on the security properties of the onion service protocol itself. Efforts to introduce application-layer encryption expose a fundamental tension: incorporating modern asynchronous protocols such as Signal would require key directories or pre-key publication, reintroducing metadata leakage and trusted infrastructure that Ricochet explicitly seeks to avoid.

Cwtch extends this design space by addressing Ricochet's lack of asynchronous messaging. Lewis~\cite{cwtch_infrastructure} proposes the use of untrusted, discardable relay servers that broadcast encrypted messages to all connected participants, enabling offline delivery without revealing sender–receiver relationships. While this approach preserves strong metadata resistance, it does so at the cost of increased bandwidth usage, broadcast semantics, and more complex availability and denial-of-service assumptions. Cwtch illustrates that limited infrastructure can coexist with strong privacy goals, but only by accepting significant performance and complexity trade-offs.

Mesh-based messengers such as Briar further generalise peer-to-peer communication by supporting multiple transports, including Bluetooth, Wi-Fi, and Tor, allowing messages to propagate opportunistically across disconnected networks. Blöchinger and von Seck~\cite{survey_mesh_messaging_2021} classify such systems as client mesh networks, where end-user devices both originate and route messages. Briar exemplifies this model by synchronising encrypted message state across contacts when connectivity becomes available, providing resilience against infrastructure failure and censorship. However, this flexibility introduces substantial state management challenges, particularly for group communication, consistency, and key evolution.

These challenges are underscored by Song's cryptographic analysis of Briar~\cite{cryptography_in_the_wild_briar}, which identifies weaknesses related to handshake forward secrecy, denial-of-service resilience, and group message duplication. This analysis illustrates a recurring pattern across serverless and mesh-based messengers: eliminating central servers often necessitates bespoke cryptographic constructions and custom protocol logic, increasing both implementation complexity and verification difficulty.

Across Ricochet, Cwtch, and Briar, common design tensions emerge. Systems that prioritise serverlessness and metadata resistance frequently sacrifice low-latency delivery, global availability, or seamless multi-device support. Discovery mechanisms are typically out-of-band, presence information is limited, and group messaging scales poorly without introducing additional infrastructure. Nonetheless, these systems demonstrate that meaningful secure communication without central servers is achievable, provided that trade-offs are made explicit and carefully bounded.

Ember draws directly from this lineage while occupying a distinct point in the design space. Rather than relying on Tor onion services or ad hoc mesh routing, Ember leverages an encrypted IPv6 overlay network to provide stable peer addressing and transport confidentiality. This approach retains the benefits of direct peer-to-peer communication while simplifying routing semantics and avoiding broadcast or store-and-forward relays. Combined with an Android-first implementation and aggressive application-level data minimisation, Ember positions itself as a pragmatic evolution of prior serverless messaging systems rather than a wholesale replacement for federated or centralised platforms.

\subsection{Decentralised IPv6 Overlay Networking and Yggdrasil as an Enabling Substrate}

A defining characteristic of Ember is its reliance on an encrypted IPv6 overlay network rather than a traditional client--server transport stack or a Tor-like anonymity system. This places the system within a small but growing body of work that explores decentralised, key-derived addressing and self-organising routing as an alternative to centrally coordinated communication infrastructures.

Early exploratory work in this space includes an engineering evaluation of Matrix federation over Yggdrasil by Floury at EPFL~\cite{matrix_yggdrasil_experiment}. Although not peer reviewed, the report provides a practical assessment of how a complex messaging protocol behaves when deployed over a cryptographically addressed overlay network. The study highlights several advantages of Yggdrasil, including automatic peering, encrypted links, stable key-derived IPv6 addressing, and improved connectivity in NAT-restricted or hostile network environments. At the same time, it explicitly notes limitations such as routing convergence delays, sensitivity to topology changes, and the absence of built-in metadata-resistance properties. These observations reinforce the view that such overlays function primarily as transport substrates rather than as comprehensive privacy systems.

A more systematic evaluation of Yggdrasil's routing properties is provided by Pestov and Kyrychek~\cite{yggdrasil_routing_scheme}. Their large-scale experiments examine scalability, routing behaviour, and resource consumption across emulated topologies ranging from tens to hundreds of nodes. The results show that Yggdrasil maintains a modest and predictable memory footprint and supports long multi-hop paths without imposing a practical hop limit, making it viable for resource-constrained devices. However, the evaluation also identifies trade-offs inherent to its spanning-tree-based routing model. Performance characteristics depend on root node placement, and topology dynamics can introduce routing inconsistencies if root selection is unstable. Proposed mitigations, such as deliberately generating and pinning a stable root node, operate at the deployment level rather than being enforced by the protocol itself.

This body of work positions Yggdrasil as a capable but deliberately limited substrate for decentralised communication. It provides encrypted point-to-point transport, globally unique and collision-resistant addressing derived from cryptographic keys, automatic peer discovery, and reasonably stable routing under semi-static conditions. However, it does not attempt to provide metadata privacy, unlinkability, cover traffic, or resistance to traffic-analysis adversaries. Any such guarantees must therefore be implemented at higher layers, if required.

For Ember, Yggdrasil functions as an enabler rather than as a security primitive. Its key-derived addressing model aligns cleanly with Ember's peer identity and binding mechanisms, and its decentralised routing eliminates the need for central servers, bootstrap infrastructure, or onion-service coordination. Ember deliberately does not treat the overlay as a source of anonymity or metadata protection: adversaries with visibility into overlay entry and exit points may still infer communication relationships and timing patterns. By explicitly framing Yggdrasil as a reachability and transport mechanism rather than a privacy enhancement, Ember adopts a conservative security posture that avoids conflating network connectivity with resistance to traffic analysis.

In summary, Yggdrasil occupies a distinct niche in the design landscape. It provides the connectivity foundation required for a serverless, mobile-first secure messenger without incurring the engineering overhead of Tor or the operational complexity of classical mesh routing protocols. Existing experimental and academic work validates its feasibility and scalability while clearly delineating its limitations. Ember builds upon this substrate to explore a middle ground between peer-to-peer security and practical deployability, acknowledging that stronger metadata-resistance mechanisms remain an area for future work.

\subsection{Ephemerality, Deletion Semantics, and Data Remanence on Android}

Ephemeral messaging features such as message timers, automatic purging of local state, and enforced deletion windows have become a prominent part of secure messaging user experience. However, extensive empirical evidence across Android versions demonstrates that logical deletion at the application layer does not imply physical secure deletion at the storage layer. For a system such as Ember, which aims to minimise retained plaintext and enforce short-lived visibility of messages, this literature imposes a fundamental constraint: TTL-based deletion can remove application-level access to content, but cannot guarantee eradication of underlying data traces under realistic forensic adversaries.

Foundational work on Android data remanence by Vidas et al.~\cite{android_data_residue_attacks_2017} demonstrates that deleted application data frequently persists across multiple system locations, including filesystem journals, app-specific directories, and system service caches. Even after application uninstallation, artefacts may remain accessible through orphaned inodes, cached credentials, or residual inter-process state. The authors show that such residues can be exploited to recover sensitive material or support follow-on attacks, directly challenging assumptions that purging application storage meaningfully eliminates forensic artefacts.

Subsequent filesystem-level analysis reinforces this conclusion. Reardon et al.~\cite{why_data_deletion_fails_2017} show that Android's deletion primitives on \texttt{ext4} do not reliably clear inode metadata, directory entries, or journaling structures. Their experiments demonstrate that substantial portions of deleted data can remain recoverable even after prolonged device usage, and that file-carving techniques can reconstruct deleted SQLite databases and chat artefacts in the absence of filesystem metadata. More recent forensic analysis by Anglano~\cite{digital_forensics_android_privacy} confirms that similar weaknesses persist on Android 9 and 10, with deleted artefacts frequently recoverable from journaling regions, write-ahead logs, and unallocated blocks despite the presence of TRIM and full-disk encryption mechanisms.

Beyond raw storage-level remanence, empirical studies of real-world messaging applications reveal additional inconsistencies introduced at the application layer. Heath et al.~\cite{forensic_ephemeral_messaging} analyse disappearing message features across WhatsApp, Telegram, and Snapchat, showing that expired content commonly persists in cache directories, quoted-message structures, forwarded-message logs, notification artefacts, and cloud backups. These findings demonstrate that ephemerality is often enforced at the user-interface or protocol level rather than through comprehensive suppression of underlying artefacts, leaving multiple residual exposure paths.

User perception further complicates this picture. Schnitzler et al.~\cite{user_perceptions_deletion} show that users frequently conflate logical deletion with secure erasure, and often assume that disappearing messages eliminate all traces by default. These assumptions can expand the effective threat surface, as users may rely on ephemerality for sensitive communication without appreciating the fact that data may persist elsewhere.

This body of work imposes strict boundaries on what Ember can plausibly claim. Ember's design deliberately minimises plaintext exposure by avoiding persistent message databases, enforcing ciphertext-only storage, aggressively clearing in-app state, and applying TTL-driven deletion of message records. This removes a substantial class of application-layer artefacts, particularly those arising from long-lived plaintext stores, message indexing structures, and notification previews. However, Ember cannot guarantee secure deletion at the level of the Android filesystem. As the cited literature demonstrates, deleted data may persist in journaling regions, write-ahead logs, cached metadata, or untrimmed flash storage, and may be recoverable by a forensically capable adversary with physical device access.

Accordingly, Ember explicitly constrains its threat model. TTL-based deletion ensures that expired messages become inaccessible through the Ember application interface, but does not assert physical sanitisation of underlying storage. Full-disk encryption and hardware-backed keystore protections mitigate some risks by binding decryption capability to user authentication and trusted execution environments. Nevertheless, ephemerality on Android must be framed as limiting live exposure and post-compromise readability \emph{within the application}, rather than as preventing all forms of forensic recovery under device compromise.

\subsection{Mobile Implementation Reality: Push Notifications, Metadata Leakage, and Android Key Storage}

While Ember avoids reliance on central servers for message relay, mobile operating systems impose architectural constraints that materially affect privacy and key-management guarantees. Two aspects are particularly relevant: metadata leakage via push notification infrastructures, and the practical realities of secure key storage on Android devices. Both represent adversarial surfaces that persist even when transport- and protocol-level security is strong.

\paragraph{Push Notifications as a Metadata Leakage Vector.}  
Recent empirical work shows that push notification infrastructures frequently undermine the privacy guarantees of otherwise secure messaging systems. Samarin et al.~\cite{push_notification_leakage} analyse 21 popular messaging applications and demonstrate that more than half transmit sensitive metadata to third-party push providers such as Google Firebase Cloud Messaging (FCM), often in plaintext. Leaked information includes user identifiers, sender and recipient details, and in some cases message content itself. Importantly, this leakage arises not from cryptographic weaknesses in end-to-end encryption protocols, but from application-layer notification payloads routed through external infrastructure.

The study highlights a systemic issue: developer-facing push notification APIs encourage embedding rich payload data, while end-to-end encryption at the notification layer is left as an optional and frequently omitted safeguard. As a result, even applications that correctly implement strong cryptographic protocols may inadvertently expose metadata through push delivery channels. These findings are reinforced by documented legal processes in which push tokens and associated metadata have been obtained from platform providers.

For Ember, this literature motivates a deliberate avoidance of third-party push notification services for message delivery. By relying on a serverless peer-to-peer overlay rather than FCM or APNs, Ember eliminates a well-documented class of metadata leakage pathways. This design choice trades OS-native wake-up mechanisms for tighter control over message transport and metadata exposure, making availability and delivery explicit responsibilities of the application rather than outsourced infrastructure.

\paragraph{Android Key Storage: Empirical Use and Misuse.}  
A second constraint concerns secure key storage on Android. While the platform exposes trusted hardware through the Keystore API, empirical evidence shows that its protections are inconsistently used in practice. Blessing et al.~\cite{android_key_storage_analysis_2025} present a large-scale measurement study analysing hundreds of thousands of Android applications and find widespread misuse or avoidance of hardware-backed key storage. A majority of apps handling sensitive data do not use trusted hardware at all, and only a small fraction leverage the StrongBox secure element. Even among applications that invoke the Keystore API, insecure parameter choices and delegation of key management to third-party libraries are common.

The study further shows that performance and latency considerations strongly influence developer behaviour. Hardware-backed operations, particularly those involving StrongBox, introduce measurable overheads that discourage their use in latency-sensitive paths. The analysis also clarifies a key limitation of Android's hardware-backed keystores: while they effectively prevent key exfiltration, they do not prevent an attacker with sufficient privileges from invoking those keys on-device.

These findings inform Ember's key-management strategy. Ember assumes the availability of TEE-backed keystore protections for long-term identity material where supported, but does not rely on StrongBox or device-level guarantees as a universal security baseline. Ephemeral session keys are short-lived by design and are not retained beyond active communication windows, reducing exposure even in the presence of imperfect platform security.

\paragraph{Implications for Ember.}  
This body of work reinforces two architectural positions adopted by Ember. First, metadata minimisation is achieved primarily by eliminating dependence on third-party push infrastructures rather than attempting to retrofit encryption onto inherently leaky delivery channels. Second, hardware-backed key storage is treated as a defence-in-depth mechanism rather than a foundational assumption. Ember pairs opportunistic use of trusted hardware with protocol- and application-level minimisation strategies, ensuring that confidentiality does not hinge on the correctness or availability of platform-specific security features.

The broader implication is that secure messaging on mobile platforms is shaped as much by operating system behaviour and developer-facing APIs as by cryptographic protocol design. Ember adopts a conservative posture within this environment by reducing reliance on external infrastructures, minimising retained state, and making its security guarantees explicit and narrowly scoped.

\section{Threat Model and Assumptions}
\label{sec:threat-model}
This section defines the adversary model, trust assumptions, security goals, and explicit non-goals for Ember. The intent is to make Ember's security claims evaluable against established threat models while avoiding overstated guarantees. Following prior SoK work, Ember distinguishes between \emph{content security} (confidentiality/integrity/authenticity of message payloads) and \emph{metadata exposure} (who talks to whom, when, and how often), and treats mobile platform realities and retained state as first-class constraints \cite{sok_secure_messaging_2015, push_notification_leakage}.

\subsection{Adversary Capabilities}
The authors assume the following adversarial capabilities.

\paragraph{Network observation (passive).}
An adversary may observe traffic on one or more network segments, including local Wi-Fi, upstream ISPs, or overlay-adjacent vantage points. Ember operates over an encrypted IPv6 overlay (Yggdrasil), which provides link encryption and key-derived addressing at the routing layer; however, Ember does not assume this yields anonymity or traffic-analysis resistance. An observer with visibility into overlay entry/exit points may still infer communication timing and peer relationships through correlation, even if payloads are encrypted \cite{metadata_privacy_beyond_tunneling_2024, yggdrasil_routing_scheme}.

\paragraph{Active network interference.}
An adversary may drop, delay, replay, and reorder packets; inject malformed frames; attempt downgrade or protocol-confusion attacks; and perform denial-of-service (DoS) against the transport layer (e.g., repeated connection attempts, resource exhaustion against the listener). Ember's application-layer envelope includes authenticated ciphertext (AES-GCM) and an explicit HMAC gate; consequently, successful payload tampering should be detected before decryption, but availability is not guaranteed under sustained DoS.

\paragraph{Malicious peers and impersonation attempts.}
A peer contact may behave arbitrarily: send malformed envelopes, attempt state desynchronisation, spam rotation requests, or attempt to impersonate a different sender identity. Ember assumes contacts are not inherently trusted. Identity is anchored to cryptographic material (conversation key fingerprints) and a pragmatic TOFU-style workflow with optional manual verification. Because Ember does not currently deploy a full asynchronous authenticated key exchange (e.g., X3DH) and does not enable Double Ratchet, it does \emph{not} claim strong resistance to key-compromise impersonation under broad conditions; instead, it prioritises correct authenticated encryption and explicit sender-binding at the message layer.

\paragraph{Bounded device compromise.}
The authors assume \emph{bounded} endpoint compromise scenarios:
\begin{itemize}
    \item \textbf{Opportunistic compromise:} malware or a local attacker gains access to the application sandbox, obtains the encrypted database, or reads screen content while the device is unlocked.
    \item \textbf{Privilege escalation / root:} a stronger attacker gains root or debugging capability, enabling memory inspection, IPC manipulation, and direct interaction with the app process.
\end{itemize}
Under root compromise, hardware-backed keystore protections may prevent key \emph{exfiltration} but do not necessarily prevent key \emph{use} on-device (i.e., an attacker can often drive cryptographic operations through the compromised process) \cite{android_key_storage_analysis_2025}. Ember therefore treats root compromise as a severe condition that can break confidentiality for currently accessible sessions and any plaintext visible in memory.

\paragraph{Physical access and forensic recovery.}
A forensically capable adversary may obtain a device image or access residual storage artefacts. Ember enforces TTL deletion and avoids plaintext persistence in the database, but the literature shows that Android deletion semantics do not imply physical erasure: artefacts can persist in journaling structures, WAL regions, and unallocated blocks \cite{why_data_deletion_fails_2017, forensic_ephemeral_messaging}. Ember explicitly limits its claims accordingly (see Non-Goals).

\subsection{Trust Assumptions}
Ember's security properties depend on explicit assumptions about endpoints, key storage, and cryptographic primitives.

\paragraph{Endpoint computing base.}
Ember assumes the endpoint OS and runtime are sufficiently trustworthy that:
\begin{itemize}
    \item the app's process and memory are not continuously instrumented by an attacker, and
    \item the UI is not systematically subverted (e.g., full accessibility abuse, screen overlay attacks) during sensitive operations such as fingerprint verification or key rotation confirmation.
\end{itemize}
This is a pragmatic assumption shared by most mobile secure messaging designs: if the endpoint is fully compromised, protocol-level guarantees degrade sharply regardless of transport properties \cite{sok_secure_messaging_2015}.

\paragraph{Android Keystore guarantees (opportunistic hardware backing).}
Ember assumes Android Keystore behaves as documented for generating and protecting long-term secret material, with hardware-backed storage (TEE) available on many devices. However, Ember does \emph{not} assume StrongBox availability or performance suitability, and does not rely on StrongBox for latency-sensitive operations. This aligns with empirical observations that secure hardware usage is fragmented and frequently avoided due to performance and integration constraints \cite{android_key_storage_analysis_2025}.

\paragraph{Cryptographic primitive security.}
Ember assumes standard cryptographic primitives are secure and correctly implemented:
AES-256-GCM for confidentiality/integrity, HMAC-SHA256 for explicit integrity gating and sender-binding, HKDF-SHA256 for key derivation, and SHA-256 for fingerprints and identifiers. Ember also assumes a cryptographically secure RNG for nonce generation and secret derivation. Under these assumptions, message confidentiality and integrity reduce to correct key management and correct use of authenticated encryption, rather than relying on the overlay network for security.

\paragraph{Overlay transport as reachability, not anonymity.}
Ember treats Yggdrasil as a reachability and encrypted-link substrate that simplifies peer connectivity via stable key-derived IPv6 addressing and encrypted links. Ember does not treat the overlay as providing sender/receiver anonymity, unlinkability, cover traffic, or resistance to global traffic analysis; these remain out of scope without additional anonymity mechanisms \cite{yggdrasil_routing_scheme, metadata_privacy_beyond_tunneling_2024}.

\subsection{Security Goals}
Ember aims to satisfy the following goals within the above assumptions and constraints.

\paragraph{Confidentiality of message content.}
Only intended endpoints should be able to decrypt message payloads. Ember enforces this via application-layer authenticated encryption (AES-256-GCM) using per-conversation symmetric keys, with plaintext never persisted to disk and decrypted only in memory for rendering. Ciphertext-only persistence is intended to minimise risk from device seizure or sandbox-level data exfiltration.

\paragraph{Integrity and authenticity of messages.}
Recipients should detect tampering and reject forged messages. Ember's envelope processing uses an explicit HMAC verification step prior to decryption (verify-then-decrypt), and binds sender metadata into the authenticated material to mitigate trivial spoofing in the absence of a full ratcheting identity system. This goal is aligned with baseline secure messaging objectives under an active network adversary \cite{sok_secure_messaging_2015}.

\paragraph{Data minimisation and bounded retention.}
Ember aims to minimise retained sensitive data by:
\begin{itemize}
    \item storing only ciphertext at rest,
    \item enforcing TTL expiration through WorkManager cleanup, and
    \item avoiding push-notification payloads that disclose content or sensitive identifiers.
\end{itemize}
This is motivated both by secure messaging threat models that treat metadata and retained artefacts as primary risks, and by empirical evidence that "secure" messengers often leak sensitive data through auxiliary infrastructures \cite{push_notification_leakage}.

\paragraph{Practical key lifecycle management.}
Ember targets a key lifecycle that is implementable on Android without excessive state complexity:
\begin{itemize}
    \item long-term protection of conversation keys via Keystore-backed storage where available,
    \item explicit user-verifiable fingerprints to support pragmatic trust establishment,
    \item key rotation to reduce exposure from long-lived keys and support recovery from suspected compromise, without requiring server coordination.
\end{itemize}
Key rotation improves resilience compared to static session keys but is not equivalent to full ratcheting-based post-compromise security (PCS).

\paragraph{Availability under realistic constraints.}
Ember aims for usable delivery and reception when peers are reachable on the overlay, but it does not claim high availability under adversarial DoS, mobile background execution limits, or prolonged peer offline periods. The system explicitly prefers minimised infrastructure over centralised reliability guarantees, consistent with its serverless design constraints.

\subsection{Explicit Non-Goals}
Ember intentionally does \emph{not} claim the following properties in its current design.

\paragraph{Anonymity and strong metadata privacy.}
Ember does not claim sender/receiver anonymity, unobservability, or robust resistance to traffic analysis against a global passive adversary. Overlay routing may reduce central collection points, but timing and relationship inference remain plausible under realistic observation models \cite{metadata_privacy_beyond_tunneling_2024}. Any stronger metadata-hiding properties would require additional mechanisms (e.g., cover traffic, mixing, rendezvous privacy), which are out of scope for the present implementation.

\paragraph{Full forward secrecy and post-compromise security at Signal level.}
Although Double Ratchet code exists in the codebase, it is disabled, and Ember does not claim continuous forward secrecy (FS) or PCS comparable to Signal-style ratcheting protocols. Ember's current design trades ratcheting complexity for implementability and correctness in a serverless peer-to-peer environment. Key rotation reduces long-lived key exposure, but tellingly does not provide epoch-by-epoch recovery semantics under compromise windows \cite{tight_security_double_ratchet_2024}.

\paragraph{Secure deletion against forensic adversaries.}
TTL expiry and database cleanup ensure that expired messages become inaccessible through Ember and are removed from application-level stores. Ember does not claim physical secure deletion on Android storage media, nor does it claim elimination of all artefacts recoverable via filesystem journaling, WAL remnants, or raw block carving \cite{why_data_deletion_fails_2017, digital_forensics_android_privacy}. Ephemerality is therefore defined as \emph{application-level access minimisation}, not guaranteed sanitisation.

\paragraph{Group messaging and multi-device consistency.}
Ember currently targets one-to-one communication. It does not claim group messaging security properties (e.g., MLS-style epoch-based membership updates, scalable FS/PCS under group churn), nor does it claim multi-device state consistency. Extending Ember to groups would require explicit re-evaluation of authentication services, delivery semantics, and group state management assumptions \cite{rfc9420_mls}.

\paragraph{Comprehensive endpoint compromise resistance.}
Ember does not claim to protect message confidentiality if an attacker has sustained control of an unlocked device, highlights UI content via overlays, or compromises the app process at runtime. As with most messengers, endpoint compromise is treated as a dominant failure mode; Ember's mitigations focus on reducing the value of stored artefacts and limiting retention, not defeating a fully compromised client.

% =========================================================

\section{System Overview}
Ember is an Android-first, peer-to-peer secure messenger that operates over the Yggdrasil IPv6 overlay network without centralised relay infrastructure. The system is intentionally conservative in its protocol surface area: it prioritises message confidentiality and integrity, minimises retained plaintext, and constrains feature scope to keep security properties analysable and implementation risk bounded.

\subsection{High-Level Architecture}
Ember follows a layered architecture that separates user interface logic, domain orchestration, cryptographic operations, persistence, and network transport. This separation is deliberate: it reduces the risk of cross-layer security coupling, improves testability, and enables reviewers to reason about security-critical paths without conflating UI behaviour with cryptographic correctness.

\paragraph{UI layer (presentation and interaction).}
The UI is implemented using Jetpack Compose and a ViewModel-driven state model. ViewModels expose immutable UI state, mediate user actions (send message, add contact, initiate key rotation), and subscribe to reactive streams of conversation and network status. The UI layer contains no cryptographic primitives and does not access raw storage directly; it delegates all security-sensitive operations to the domain layer.

\paragraph{Domain layer (orchestration and state machines).}
The domain layer centres on a repository-style orchestration component (\texttt{ChatRepository}). This layer coordinates:
(i) encryption and authentication, (ii) persistence and retrieval of message records, (iii) transport send/receive flows, and (iv) higher-level protocols such as key rotation. The repository enforces security invariants, including \emph{verify-then-decrypt} processing, plaintext non-persistence, TTL scheduling, and key versioning rules.

\paragraph{Cryptographic layer (encryption, authentication, key management).}
Ember's active cryptographic design uses per-conversation symmetric keys with authenticated encryption for message payload confidentiality and integrity. Message encryption uses AES-256-GCM with a fresh 12-byte random nonce per message. In addition, Ember computes an explicit HMAC-SHA256 over envelope components (including sender binding fields) and verifies the HMAC prior to attempting decryption. Conversation keys are protected using Android Keystore-backed mechanisms where available, and user-visible fingerprints are derived using SHA-256 to support out-of-band verification.

Double Ratchet logic exists in the codebase but is not enabled; Ember therefore does not rely on ratcheting state machines for correctness in the current system. Instead, Ember provides an explicit key rotation workflow that evolves the per-conversation key using HKDF and versioned derivation inputs while preserving the ability to decrypt historical ciphertext.

\paragraph{Storage layer (encrypted persistence with plaintext minimisation).}
Ember persists application state using Room with SQLCipher-backed encrypted SQLite storage. The storage model is structured around contacts, conversations, and message records. The critical storage invariant is that \textbf{only ciphertext is stored} for message bodies; plaintext is computed only in-memory for rendering and is not written to disk. Messages are associated with TTL policies and an \texttt{expiresAt} timestamp, enabling automated deletion of expired records. Database schema changes are managed through explicit migrations to avoid destructive upgrade paths.

\paragraph{Network layer (serverless transport over Yggdrasil).}
Ember uses direct TCP connections between peers on Yggdrasil IPv6 addresses. Transport is realised via:
(i) an outbound client that establishes TCP connections and sends framed message envelopes, and (ii) an inbound foreground listener service that accepts connections and parses envelopes. A lightweight, length-prefixed framing format is used to delimit JSON-serialised envelopes over TCP. A bridging component forwards received envelopes from the Android service context to the dependency-injected domain layer, where integrity verification and decryption occur.

\paragraph{End-to-end message flow.}
The end-to-end pipeline is designed so that security checks occur before any plaintext is produced:
\begin{enumerate}
    \item On send, the UI requests a send operation from the domain layer.
    \item The domain layer encrypts plaintext into ciphertext (AES-GCM), computes a HMAC over the envelope fields, and persists the ciphertext record.
    \item The network layer transmits a framed envelope containing ciphertext, nonce, and authentication data.
    \item On receive, the listener parses an envelope and forwards it to the domain layer.
    \item The domain layer loads the correct conversation key, verifies the HMAC (\emph{verify-then-decrypt}), decrypts the ciphertext, and emits plaintext only to the in-memory UI projection.
    \item TTL metadata is used to schedule expiration cleanup for stored ciphertext.
\end{enumerate}
This ordering is intentional: it ensures that malformed or unauthenticated envelopes are rejected without producing plaintext, and that stored message content remains encrypted at rest.

\subsection{Design Principles}
Ember's design is guided by a small set of explicit principles that trade breadth of features for reduced trusted infrastructure and improved analysability.

\paragraph{Serverlessness and reduced infrastructure trust.}
Ember is designed to operate without central servers for identity, message relay, or account provisioning. Peers communicate directly over the overlay using Yggdrasil IPv6 addresses and a configurable port. This reduces reliance on third-party relays and avoids the typical "provider as metadata aggregation point" characteristic of centralised messaging platforms. The system explicitly treats the overlay as a reachability substrate rather than an anonymity system; Ember does not claim comprehensive metadata protection beyond minimising retained application-level data.

\paragraph{Minimal retained plaintext.}
A core objective is to minimise plaintext persistence and reduce forensic value of stored state. Ember enforces ciphertext-only message storage and performs decryption only transiently for user display. Notifications are privacy-first and do not disclose message content. TTL policies limit how long ciphertext records remain in the local database, reducing long-term retention by default.

\paragraph{Fail-closed processing and defensive parsing.}
Ember's message processing pipeline is designed to fail closed. Envelopes are authenticated before decryption, and malformed inputs are rejected early. This principle aims to bound the impact of active network adversaries and malicious peers by making "unauthenticated plaintext production" structurally difficult.

\paragraph{Constrained protocol surface for analysability.}
Ember intentionally avoids complex protocol state machines as baseline requirements. Although ratcheting protocols provide stronger compromise-resilience properties, they introduce subtle state synchronisation and verification challenges, especially in decentralised settings. Ember's current security posture instead emphasises correctness and transparency: per-conversation keys, explicit authentication gates, and a controlled key rotation mechanism that can be reasoned about and tested in isolation.

\paragraph{Pragmatic key lifecycle management.}
Rather than assuming ideal hardware security, Ember uses platform keystore protections opportunistically and pairs them with protocol-level minimisation. Key rotation is treated as a first-class workflow to reduce the exposure window of long-lived keys without requiring an online trusted service. User-visible fingerprints provide a human-verifiable trust anchor to support manual verification and mitigate silent substitution risks in adversarial environments.

\subsection{Functional Scope}
Ember's implemented functionality is intentionally constrained to a clear baseline set of capabilities. The scope below reflects the current system behaviour rather than aspirational features.

\paragraph{Implemented capabilities.}
\begin{itemize}
    \item \textbf{One-to-one (1:1) messaging:} direct peer messaging between two endpoints, with conversation identifiers deterministically derived to provide stable mapping between contacts and threads.
    \item \textbf{End-to-end payload protection:} AES-256-GCM encryption per message using per-conversation symmetric keys, with explicit HMAC-SHA256 verification prior to decryption.
    \item \textbf{Encrypted local persistence:} SQLCipher-encrypted database via Room, storing message ciphertext and metadata while ensuring plaintext is never written to disk.
    \item \textbf{TTL expiration and background cleanup:} messages carry TTL metadata and an expiration timestamp; WorkManager-based cleanup removes expired ciphertext records to enforce bounded retention even when the app is backgrounded.
    \item \textbf{Key fingerprinting and verification support:} SHA-256-derived fingerprints rendered for user comparison to support pragmatic trust establishment and verification workflows.
    \item \textbf{Key rotation:} a versioned, HKDF-based key evolution workflow that updates the per-conversation key through explicit protocol stages and preserves the ability to decrypt historical messages.
    \item \textbf{Privacy-first notifications:} notifications avoid displaying message content and use conservative lock-screen visibility settings to reduce inadvertent disclosure.
    \item \textbf{Serverless transport over Yggdrasil:} direct TCP send/receive over overlay IPv6 addresses, including an inbound listener running as a foreground service for reliable reception.
    \item \textbf{Operational diagnostics:} UI-accessible network and status views intended to support troubleshooting and to make transport state visible to the user.
\end{itemize}

\paragraph{Explicit exclusions in the current system.}
\begin{itemize}
    \item \textbf{Group messaging:} not implemented; no MLS-style group state or sender-key semantics are provided.
    \item \textbf{Double Ratchet operation:} ratcheting code exists but is disabled; Ember therefore does not claim Signal-like forward secrecy or post-compromise security in the current implementation.
    \item \textbf{Centralised identity services:} Ember does not provide accounts, phone-number identity binding, or a global directory; contacts are created through direct endpoint configuration and verification workflows.
\end{itemize}

These choices position Ember as a deliberately minimal secure messenger: it provides strong content confidentiality and integrity within a serverless peer-to-peer environment, while keeping protocol complexity low enough that the system's guarantees, failure modes, and trust dependencies remain explicit and reviewable.

\section{Network Architecture}
Ember's network architecture prioritises direct, serverless reachability and protocol simplicity over the abstraction and reliability guarantees typically offered by centralised messaging infrastructure.

\subsection{Transport Model}
Ember uses direct peer-to-peer TCP communication over an encrypted IPv6 overlay network. Each peer is addressed using a stable IPv6 address derived from cryptographic key material by the overlay, and listens on a configurable TCP port. This addressing model provides globally unique, collision-resistant endpoints without requiring DNS, NAT traversal services, or central rendezvous servers.

TCP is used as the transport protocol for three primary reasons. First, it provides reliable, ordered delivery semantics that simplify message handling at the application layer. Second, it allows Ember to avoid implementing its own congestion control, retransmission logic, or fragmentation scheme. Third, TCP is well-supported by Android's networking APIs and integrates cleanly with long-lived background services when required.

The overlay network provides encrypted links between nodes, but Ember does not treat the overlay as a security boundary. All application-level security properties—confidentiality, integrity, and sender authentication—are enforced independently at the message layer. As a result, Ember remains robust even if the overlay is misconfigured, partially observable, or compromised at the routing level. The overlay is therefore treated strictly as a reachability substrate rather than a trust anchor.

Communication is symmetric: every node can both initiate outbound connections and accept inbound connections. There is no concept of a relay or broker within Ember itself. Message delivery succeeds only when peers are simultaneously reachable on the overlay and listening on the expected port, reinforcing Ember's peer-to-peer and availability-bounded design philosophy.

\subsection{Envelope and Framing Protocol}
Messages are transmitted as discrete envelopes over TCP using a simple length-prefixed framing protocol. Framing is required because TCP is a byte-stream protocol with no inherent message boundaries.

Each transmitted message consists of:
\begin{enumerate}
    \item A fixed-size 4-byte big-endian unsigned integer indicating the length of the envelope payload in bytes.
    \item A JSON-serialised envelope of the specified length.
\end{enumerate}

The length prefix allows the receiver to deterministically read exactly one envelope from the stream, reject oversized or malformed payloads early, and avoid ambiguity arising from partial reads or stream concatenation. This framing strategy is intentionally minimal and avoids complex multiplexing or stream management logic.

The envelope itself contains all metadata required for message processing and verification. Typical fields include:
\begin{itemize}
    \item protocol version identifier,
    \item conversation identifier,
    \item sender identifier or name,
    \item timestamp and TTL metadata,
    \item message type (e.g.\ user message, key rotation control message),
    \item ciphertext (Base64-encoded),
    \item nonce (Base64-encoded),
    \item HMAC (Base64-encoded).
\end{itemize}

No plaintext message content is ever included in the envelope. Metadata fields are intentionally constrained to those required for routing, authentication, and lifecycle management. In particular, Ember avoids embedding user profile data, contact lists, or message previews in envelopes, reducing the risk of metadata amplification at the transport layer.

On reception, envelopes are parsed and validated defensively. Length mismatches, malformed JSON, missing required fields, or unsupported protocol versions result in immediate rejection. Crucially, cryptographic verification (HMAC validation) is performed before any decryption attempt. This ordering enforces a fail-closed processing model in which unauthenticated data cannot cause plaintext to be produced or persisted.

\subsection{Connection Management}
Connection management in Ember is intentionally simple and conservative. Each outbound message send establishes a TCP connection to the peer's IPv6 address and port using a bounded retry strategy. Long-lived connections are not assumed to be stable, reflecting the realities of mobile networking, background execution limits, and dynamic overlay routing.

Outbound connections use explicit timeouts to bound resource consumption:
\begin{itemize}
    \item connection establishment timeout,
    \item read timeout for framed envelope reception,
    \item write timeout for outbound payload transmission.
\end{itemize}

If a connection attempt fails, Ember retries using exponential backoff up to a fixed maximum number of attempts. This limits the impact of transient network failures while avoiding unbounded retry loops that could drain battery or create denial-of-service conditions.

Inbound connections are accepted by a dedicated listener service (discussed below) and processed sequentially. Ember does not attempt to multiplex multiple concurrent message streams over a single connection, nor does it maintain per-peer connection pools. This choice reduces state complexity and failure modes at the cost of some efficiency, which is acceptable given Ember's low expected message throughput and emphasis on analysability.

Connection state is modelled explicitly and exposed to the UI via reactive state streams. States such as \texttt{DISCONNECTED}, \texttt{LISTENING}, \texttt{CONNECTED}, and \texttt{ERROR} allow both users and developers to observe network behaviour directly. This transparency is intentional: rather than hiding connectivity issues behind opaque retry logic, Ember surfaces transport state to support debugging, user understanding, and security evaluation.

\subsection{Background Receive Capability}
Receiving inbound connections on Android imposes strict constraints due to modern background execution limits. Android aggressively restricts background network access and process lifetime in order to preserve battery life, making traditional always-on listeners infeasible for most applications.

To ensure reliable inbound message reception, Ember employs a foreground service that hosts the TCP listener. Running as a foreground service grants the application elevated execution priority and permits continuous socket listening while the device is idle or the UI is not in the foreground. This design choice is explicit and user-visible: the service presents a persistent notification indicating that Ember is actively listening for incoming connections.

The use of a foreground service is justified on functional and security grounds. Functionally, it is the only reliable mechanism to support serverless peer-to-peer reception without relying on third-party push notification infrastructure. From a security perspective, it avoids the need to route message metadata or wake-up signals through external providers, which have been shown to introduce significant privacy leakage in mainstream messaging systems.

However, this approach comes with clear trade-offs. Foreground services consume more system resources, may impact battery life, and require explicit user consent. Ember treats these costs as acceptable within its threat model and design goals, preferring transparent, user-controlled availability over opaque reliance on centralised notification channels.

In summary, Ember's network architecture is deliberately straightforward: direct TCP over cryptographically addressed IPv6 peers, minimal framing and envelope semantics, conservative connection management, and explicit foreground execution for receive capability. This simplicity reduces the attack surface, makes failure modes explicit, and aligns with Ember's broader goal of providing secure peer-to-peer messaging without hidden infrastructure dependencies.

% =========================================================

\section{Cryptographic Design}
Ember's cryptographic design deliberately prioritises correctness, explicit security boundaries, and ease of analysis over maximal theoretical guarantees. The system enforces strong confidentiality and integrity for message payloads while avoiding complex asynchronous ratcheting state as a baseline requirement.

\subsection{Conversation Identity and Keying}
Each Ember conversation is defined as a one-to-one relationship between two peer endpoints. Conversations are identified using a deterministic conversation identifier derived from both peers' addressing material. The identifier is computed as a cryptographic hash over a canonicalised representation of the two endpoints (e.g.\ IPv6 address and port), ensuring that:
\begin{itemize}
    \item the same pair of peers always derives the same conversation identifier,
    \item conversation identifiers are collision-resistant under standard hash assumptions,
    \item no external coordination or server-side allocation is required.
\end{itemize}

For each conversation, Ember maintains a single symmetric conversation key that is shared between the two peers. This key is used exclusively for message payload encryption and authentication within that conversation. Keys are scoped strictly to the conversation context: compromise of one conversation key does not affect other conversations, even with the same peer.

The conversation key lifecycle is explicit. Keys are created during contact establishment, referenced by versioned identifiers, and may be evolved using Ember's key rotation protocol. Historical keys are retained only to the extent necessary to decrypt previously stored ciphertext, allowing forward progression without invalidating message history.

\subsection{Confidentiality Mechanism}
Message confidentiality is provided using AES-256-GCM, an authenticated encryption with associated data (AEAD) construction that provides confidentiality and integrity under standard cryptographic assumptions. Ember uses a 256-bit symmetric key per conversation and generates a fresh, cryptographically secure random nonce for each message.

Nonce generation follows a strict uniqueness requirement: a 12-byte nonce is generated using a cryptographically secure random number generator for every encryption operation. Nonces are never reused with the same conversation key. This design avoids catastrophic failure modes associated with nonce reuse in GCM and does not rely on counters or shared state synchronisation between peers.

AES-GCM is used solely for encrypting message payloads. Plaintext exists only transiently in memory during encryption and decryption operations and is never persisted to disk. Ciphertext, nonce, and authentication data are persisted as part of the message record, enabling deferred decryption without exposing plaintext at rest.

Under these conditions, Ember provides semantic security against passive network adversaries and confidentiality of message content against untrusted intermediaries, provided that the conversation key remains secret and nonce uniqueness is preserved.

\subsection{Message Integrity and Authenticity}
While AES-GCM already provides integrated authentication, Ember introduces an explicit HMAC-SHA256 layer as a separate integrity and authenticity gate. This design choice serves three purposes:
\begin{enumerate}
    \item enforcing a strict \emph{verify-before-decrypt} processing model,
    \item explicitly binding sender metadata into the authenticated material,
    \item simplifying reasoning about envelope-level integrity independent of the AEAD primitive.
\end{enumerate}

For each message, Ember computes an HMAC-SHA256 over selected envelope fields, including at minimum the ciphertext and sender-identifying metadata. The HMAC key is derived from the conversation key. On reception, the HMAC is verified prior to any decryption attempt. Messages with invalid or missing HMACs are rejected immediately and never decrypted.

This ordering is intentional. By authenticating data before decryption, Ember prevents unauthenticated ciphertext from influencing decryption logic or producing plaintext output, reducing the risk of cryptographic misuse and malformed-input exploitation. Sender binding further mitigates trivial spoofing attacks in which a malicious peer attempts to inject messages while claiming a different sender identity.

The explicit HMAC layer does not aim to replace full identity authentication as provided by ratcheting protocols with signed key exchanges. Instead, it provides a clear, auditable integrity boundary that aligns with Ember's constrained threat model and implementation scope.

\subsection{Key Storage}
Long-term conversation keys are protected using Android's Keystore framework, accessed via the Android Security Crypto library. Where available, keys are backed by hardware-enforced Trusted Execution Environment (TEE) protections, preventing raw key material from being exported from secure storage.

Ember treats Keystore-backed storage as an opportunistic defence rather than an absolute guarantee. Hardware-backed storage can significantly raise the bar for key exfiltration by malware or sandbox-level attackers, but it does not prevent all attack classes. In particular:
\begin{itemize}
    \item a compromised application process may still invoke cryptographic operations using stored keys,
    \item root-level adversaries may manipulate runtime behaviour even if key bytes remain non-exportable,
    \item StrongBox-backed storage is not assumed due to limited device support and high performance overhead.
\end{itemize}

Ephemeral message keys derived during encryption and decryption are not persisted beyond their immediate use. By minimising long-lived secret material and pairing Keystore protection with short-lived session data, Ember reduces its dependence on platform guarantees that are known to vary widely across devices.

\subsection{Key Fingerprinting and Verification}
To support pragmatic trust establishment, Ember derives a user-visible fingerprint for each conversation key. The fingerprint is computed as a SHA-256 hash of the raw conversation key material, with a fixed-length prefix rendered in a human-readable format (e.g.\ colon-separated hexadecimal bytes).

The fingerprint serves two purposes. First, it provides a stable identifier that users can compare out-of-band (e.g.\ verbally or via a secondary channel) to detect key substitution attacks. Second, it anchors Ember's trust model in a transparent, inspectable artifact rather than implicit assumptions about network or overlay security.

Ember follows a TOFU-style verification model by default: the first observed key for a conversation is accepted and displayed to the user, and subsequent changes are surfaced explicitly. Key rotation events update the fingerprint in a controlled manner, allowing users to distinguish legitimate key evolution from unexpected key replacement.

This approach does not provide the same formal guarantees as certificate-based public key infrastructures or fully authenticated asynchronous handshakes. However, it aligns with Ember's serverless design goals and avoids introducing central trust anchors, while still enabling users to detect and respond to suspicious key changes in adversarial environments.

% =========================================================

\section{Key Lifecycle Management}
Ember explicitly separates \emph{key lifecycle management} from advanced ratcheting-based protocols, treating key evolution as a controlled, auditable process that improves security over static long-lived keys without incurring the complexity and fragility of asynchronous ratcheting in a serverless environment.

\subsection{Key Establishment Model}
Ember's key establishment model is deliberately simple and assumes an explicit onboarding step between peers. Unlike centralised messengers that rely on directory services, phone-number binding, or pre-key servers, Ember does not provide an automated global key discovery or asynchronous authenticated key exchange.

Initial key establishment occurs during contact onboarding. When a user adds a contact, the peer's endpoint information (overlay IPv6 address and port) is configured locally, and a fresh symmetric conversation key is generated for that conversation. This key becomes the root secret for all subsequent message encryption and authentication within the conversation.

Trust establishment follows a pragmatic TOFU-style model. The first key observed for a conversation is accepted as authoritative and displayed to the user via a fingerprint. Ember assumes that the initial exchange occurs in a context where active man-in-the-middle interference is either unlikely or can be detected through out-of-band verification. This assumption mirrors that of many decentralised and peer-to-peer systems that operate without a public key infrastructure.

Importantly, Ember does not assume that peers are continuously online during onboarding, nor does it depend on asynchronous pre-key publication. This avoids introducing server-side coordination points or persistent public key directories, at the cost of placing greater responsibility on users to verify peer identities when security is critical.

\subsection{Key Rotation Protocol}
To reduce exposure from long-lived symmetric keys, Ember implements an explicit key rotation protocol that allows both peers to agree on a new conversation key while retaining access to historical ciphertext.

The protocol is structured as a small, finite-state exchange between peers, with clearly defined stages:
\begin{enumerate}
    \item \textbf{RotationRequest:} one peer initiates rotation by proposing a new key version identifier.
    \item \textbf{RotationConfirm:} the receiving peer acknowledges the request and confirms readiness to rotate.
    \item \textbf{RotationActivate:} both peers locally derive and activate the new key version.
\end{enumerate}

Key derivation uses HKDF with SHA-256 to deterministically evolve the next conversation key from the current one. The derivation incorporates explicit context strings and version identifiers:
\begin{itemize}
    \item the \texttt{salt} includes a fixed protocol label and the current key version,
    \item the \texttt{info} field encodes the target version identifier.
\end{itemize}
This construction ensures domain separation between key versions and prevents accidental key reuse across different protocol stages.

Once rotation is activated, new outbound messages are encrypted using the new key version. Older key versions are retained locally for the sole purpose of decrypting historical ciphertext already stored in the database. Ember does not re-encrypt past messages, avoiding unnecessary computational overhead and reducing the risk of data loss.

Mutual confirmation is a critical property of the protocol. Rotation only completes when both peers have explicitly acknowledged the transition, preventing unilateral key changes that could silently desynchronise conversation state or cause message loss. Rotation state is tracked at the domain layer and surfaced to the UI, making key evolution visible rather than implicit.

\subsection{Security Implications of Rotation}
Key rotation improves Ember's security posture relative to static per-conversation keys, but it is not equivalent to ratcheting-based designs such as the Double Ratchet.

The primary security benefit of rotation is reduction of the \emph{key exposure window}. If a conversation key is compromised at some point in time, subsequent rotation limits the amount of future traffic that remains decryptable under the compromised key. This is particularly valuable in environments where long-lived mobile keys may be exposed through backup leakage, debugging artifacts, or partial device compromise.

However, Ember's rotation model has clear limitations. Because rotation is explicit and infrequent relative to message volume, it does not provide per-message forward secrecy. Compromise of the active key allows decryption of all messages encrypted under that key version until rotation occurs. Additionally, because historical keys are retained to preserve message history, compromise of local storage may still expose past ciphertext to offline attack, albeit protected by strong symmetric encryption.

Unlike ratcheting protocols, Ember's rotation does not provide post-compromise security (PCS) in the strong sense. If an attacker maintains persistent access to a device or application process, they can continue to observe or influence key material even after rotation. Recovery from such compromise requires explicit remediation, such as re-onboarding the contact or re-establishing trust through a fresh key.

These limitations are intentional and explicitly acknowledged. Ember treats key rotation as a pragmatic middle ground: it meaningfully improves security over static keys, is implementable without central infrastructure, and remains transparent and analysable. Advanced ratcheting-based guarantees are treated as future work rather than baseline assumptions, avoiding overclaiming security properties that are not robustly supported by the current architecture.

\section{Data Persistence and Ephemerality}
Ember's storage design is informed by empirical work showing that mobile storage layers do not provide reliable secure deletion guarantees. The system treats ephemerality as an \emph{application-level access control property}, not as a claim of physical data erasure.

\subsection{Encrypted Storage Model}
Ember persists local application state using an encrypted SQLite database implemented via Room with SQLCipher. The database provides confidentiality for stored records against sandbox-level attackers and opportunistic data extraction, assuming the encryption key remains protected by the operating system and keystore mechanisms.

The database schema is intentionally minimal and structured around three core entities:
\begin{itemize}
    \item \textbf{Contacts:} peer endpoint metadata, including display name, overlay address, port, verification state, and references to associated conversation keys.
    \item \textbf{Conversations:} one-to-one conversation metadata, including a deterministic conversation identifier, last-activity timestamp, and default TTL policy.
    \item \textbf{Messages:} per-message records containing ciphertext, cryptographic metadata (nonce, authentication tag or HMAC), sender identifier, TTL parameters, and expiration timestamp.
\end{itemize}

No application logs, message previews, or derived plaintext artefacts are stored in the database. Schema evolution is handled via explicit migrations rather than destructive resets, ensuring that encrypted data is not silently discarded or rewritten during application upgrades.

\subsection{Ciphertext-Only Persistence}
A central invariant in Ember's storage design is that \textbf{plaintext message content is never written to persistent storage}. Plaintext exists only transiently in memory during encryption (send path) and decryption (receive and render path).

On the send path, plaintext entered by the user is immediately encrypted using the active conversation key. Only the resulting ciphertext and associated cryptographic metadata are persisted. On the receive path, ciphertext is stored first and decrypted only when required for UI rendering. Decrypted plaintext is held in memory-bound data structures and discarded when no longer needed for display.

This separation ensures that:
\begin{itemize}
    \item disk-level compromise of the application database yields only encrypted payloads,
    \item message history at rest remains protected even if TTL cleanup is delayed,
    \item no plaintext artefacts exist in long-lived storage structures such as SQLite tables, WAL files, or backup images created by the application.
\end{itemize}

While in-memory plaintext is still vulnerable to runtime compromise or memory inspection, confining plaintext to volatile memory materially reduces the forensic value of persistent storage and aligns with Ember's data minimisation goals.

\subsection{TTL Expiration and Cleanup}
Ember enforces ephemerality through explicit time-to-live (TTL) policies associated with each message. Each message record includes a TTL value and a computed \texttt{expiresAt} timestamp. Once this timestamp has passed, the message is considered expired and eligible for deletion.

Cleanup is implemented using Android's WorkManager framework, which provides two complementary mechanisms:
\begin{itemize}
    \item \textbf{Periodic sweeps:} a recurring cleanup task runs at the minimum interval permitted by the platform (currently 15 minutes), deleting all messages whose expiration timestamp has elapsed.
    \item \textbf{Targeted cleanup jobs:} where supported, delayed tasks are scheduled at message expiry time to tighten deletion latency.
\end{itemize}

This design reflects practical platform constraints. Android does not guarantee precise execution timing for background work, especially under power-saving modes or aggressive OEM-specific process management. As a result, TTL expiration is enforced on a best-effort basis within bounded delay, rather than at exact wall-clock expiry.

Deletion semantics are explicitly scoped. When a message expires, its ciphertext record and associated metadata are removed from the application database, making the message inaccessible through the Ember interface. Ember does not claim that this deletion securely erases underlying storage blocks or filesystem artefacts. As documented in the literature, deleted data may persist in journaling regions, write-ahead logs, or unallocated blocks even after application-level removal.

Accordingly, Ember frames ephemerality as limiting \emph{continued access and exposure} rather than guaranteeing physical sanitisation. TTL expiration ensures that expired messages are not displayed, indexed, or retained by the application, and that long-term plaintext availability is minimised. Protection against forensic recovery beyond this point relies on full-disk encryption and hardware-backed key protection rather than on application-level deletion semantics alone.

% =========================================================
\section{Message Processing Pipelines}
Ember's message processing pipelines are structured to enforce security invariants---particularly authentication before decryption and plaintext minimisation---while remaining transparent and straightforward to reason about.

\subsection{Sending Pipeline}
The sending pipeline begins at the user interface and progresses through the domain, cryptographic, storage, and network layers in a strictly ordered manner.

When a user submits a message, the UI layer forwards the plaintext input to the domain layer via the appropriate ViewModel. The UI itself performs no cryptographic operations and does not interact with persistent storage. This ensures that security-sensitive logic is centralised and consistent.

Within the domain layer, the repository retrieves the active conversation context, including the current conversation key and its version. The plaintext message is immediately passed to the cryptographic layer, where it is encrypted using AES-256-GCM with a freshly generated random nonce. An HMAC-SHA256 is then computed over the relevant envelope fields to provide explicit integrity protection and sender binding.

Once cryptographic processing completes, the resulting ciphertext, nonce, HMAC, and associated metadata (conversation identifier, sender identifier, TTL parameters, and timestamp) are persisted to the encrypted local database. Plaintext is not written to disk at any point in this process. TTL cleanup tasks are scheduled based on the message's expiration time.

After persistence, the domain layer constructs a network envelope and hands it to the network layer for transmission. The network client establishes a TCP connection to the peer's overlay address, frames the envelope using length-prefixed encoding, and transmits the payload. Transmission is asynchronous with respect to the UI, allowing the user interface to remain responsive even if network delivery is delayed or retried.

\subsection{Receiving Pipeline}
The receiving pipeline is designed to ensure that unauthenticated or malformed data cannot result in plaintext production or persistent side effects.

Inbound TCP connections are accepted by a dedicated listener service running in the foreground. The service reads the length prefix and corresponding envelope bytes from the socket, enforcing basic size and framing validation. Parsed envelopes are then forwarded through a controlled bridge into the domain layer, where all security-sensitive processing occurs.

Upon receipt in the domain layer, the repository performs a series of checks in a fixed order:
\begin{enumerate}
    \item \textbf{Structural validation:} ensure required fields are present and the protocol version is supported.
    \item \textbf{Conversation resolution:} map the envelope to an existing conversation context using the deterministic conversation identifier.
    \item \textbf{Integrity verification:} compute and verify the HMAC using the appropriate conversation key before attempting decryption.
\end{enumerate}

Only if integrity verification succeeds does the system proceed to decryption. The ciphertext is decrypted using AES-256-GCM and the associated nonce. If decryption fails, the message is discarded without producing plaintext.

Following successful decryption, the ciphertext record and associated metadata are persisted to the encrypted database. Decrypted plaintext is retained only in memory for immediate processing. TTL metadata is recorded and cleanup scheduling is updated accordingly.

Finally, a privacy-first notification is generated to inform the user of the new message. Notifications intentionally omit message content and sensitive identifiers, ensuring that receipt alerts do not undermine the confidentiality guarantees of the encrypted payload.

\subsection{Message Display Pipeline}
Message display is decoupled from message storage and network reception. The UI observes reactive data streams that emit message records as they become available or are updated.

When a conversation view is rendered, the domain layer retrieves the relevant ciphertext records from the database and decrypts them on demand. Decryption occurs in-memory and only for messages that are actively being displayed. Plaintext is exposed to the UI as ephemeral view models and is not written back to persistent storage or cached beyond the lifetime of the UI session.

This design ensures that:
\begin{itemize}
    \item plaintext exists only transiently and only when required for rendering,
    \item expired messages that have been removed from storage cannot be redisplayed,
    \item background processes and notification handlers never access plaintext message bodies.
\end{itemize}

When the user navigates away from a conversation or the UI process is suspended, in-memory plaintext objects are eligible for garbage collection. By confining plaintext handling to the display pipeline and enforcing strict boundaries between storage and rendering, Ember minimises the risk of unintended plaintext retention while maintaining a responsive user experience.

% =========================================================

\section{Privacy and Metadata Considerations}
Consistent with prior secure messaging research, Ember treats metadata as a first-class concern but adopts a deliberately bounded position: it aims to \emph{avoid unnecessary metadata leakage} and \emph{minimise retained artefacts} rather than claiming comprehensive anonymity or traffic-analysis resistance.

\subsection{Notification Privacy}
Notifications are a common source of inadvertent metadata and content leakage in mobile messaging systems. Ember adopts a content-free notification model to ensure that message receipt alerts do not undermine end-to-end confidentiality.

When a new message is received, Ember generates a generic notification indicating only that a message has arrived. Message bodies, sender names, conversation identifiers, and previews are intentionally omitted. Notifications use conservative lock-screen visibility settings, ensuring that even the presence of a notification reveals minimal information when the device is locked.

This design avoids two well-documented classes of leakage. First, it prevents plaintext or partially decrypted message content from being exposed through the operating system's notification subsystem. Second, it avoids transmitting sensitive metadata through third-party push notification providers, since Ember does not rely on external push infrastructure for message delivery or wake-up signalling. Notifications are generated locally after authenticated message reception, preserving the confidentiality boundary enforced by the cryptographic layer.

\subsection{Metadata Exposure Analysis}
Ember's architecture avoids several common sources of metadata exposure while accepting others as unavoidable under its threat model.

\paragraph{Avoided metadata exposure.}
By design, Ember avoids:
\begin{itemize}
    \item \textbf{Centralised server metadata:} there is no server-side message relay, account system, or global directory capable of observing social graphs or message timing across users.
    \item \textbf{Push notification leakage:} message delivery and wake-up do not rely on third-party notification services that may receive sender, recipient, or content metadata.
    \item \textbf{Persistent plaintext artefacts:} message content is never stored in plaintext on disk, and expired messages are removed from application storage.
    \item \textbf{Application-level logging of sensitive events:} Ember avoids logging message content or cryptographic material, reducing the risk of leakage through debug logs or crash reports.
\end{itemize}

\paragraph{Unavoidable metadata exposure.}
Some metadata exposure remains inherent to Ember's design:
\begin{itemize}
    \item \textbf{Network-level observability:} peers communicate directly over an IPv6 overlay. An observer with visibility into overlay entry or exit points can potentially infer communication relationships, timing, and frequency.
    \item \textbf{Endpoint addressing:} stable overlay addresses are required for reachability. While key-derived addressing avoids central allocation, it still enables linkability of traffic to a given endpoint over time.
    \item \textbf{Local device metadata:} contact names, conversation ordering, and last-activity timestamps exist within the application state and may be observable under device compromise.
\end{itemize}

Ember treats these exposures as explicit trade-offs rather than accidental leakage. The system avoids amplifying metadata beyond what is necessary for peer-to-peer operation, but does not attempt to conceal all observable properties of communication.

\subsection{Residual Privacy Risks}
Despite Ember's mitigations, several residual privacy risks remain.

\paragraph{Network traffic analysis.}
Ember does not provide cover traffic, mixing, or anonymity guarantees. A sufficiently capable adversary observing network traffic over time may infer social relationships, message frequency, or usage patterns, even though payloads are encrypted. These risks are inherent to direct peer-to-peer communication without additional anonymity layers and are explicitly out of scope for the current system.

\paragraph{Endpoint compromise.}
If a device is compromised at runtime—through malware, root access, or physical access while unlocked—plaintext messages displayed on screen or resident in memory may be exposed. While Ember minimises stored plaintext and relies on hardware-backed key protection where available, it cannot defend against an attacker with sustained control of the endpoint. This limitation is shared by virtually all mobile messaging applications.

\paragraph{User-mediated leakage.}
As with any messaging system, users may voluntarily or inadvertently leak content through screenshots, message forwarding, or sharing outside the application. Ember does not attempt to restrict such behaviour at the operating system level.

In summary, Ember's privacy posture is intentionally conservative and explicit. The system meaningfully reduces several high-impact metadata leakage vectors—particularly those arising from centralised infrastructure and notification services—while acknowledging the remaining risks associated with direct peer-to-peer networking and endpoint security. By clearly articulating these boundaries, Ember avoids overstating its guarantees and allows its privacy properties to be evaluated realistically against its stated threat model.

% =========================================================

\section{Security Testing and Verification}
\label{sec:security-testing}

This section reports the results of a systematic security assessment of Ember 4.0, combining (i) static code analysis, (ii) unit-test based cryptographic validation and micro-benchmarking, (iii) OWASP ASVS v4.0 requirement mapping, (iv) structured threat modelling with mitigation-to-risk analysis, and (v) dynamic network security analysis using on-the-wire packet capture correlated with device-level runtime telemetry. The objective is to provide publication-grade evidence for Ember's implemented security properties, while explicitly documenting evaluation constraints and residual gaps.

\subsection{Methodology and Scope}
\label{subsec:sec-test-method}

The assessment was conducted in two phases. The first phase (static) comprised ruleset-driven code scanning, unit-test validation, ASVS mapping, and structured threat modelling. The second phase (dynamic) introduced on-the-wire packet capture and correlated Android runtime telemetry to validate transport and application-layer encryption behaviour under live operating conditions.

\begin{itemize}
    \item \textbf{Static analysis:} ruleset-driven scanning across the Android/Kotlin codebase, focusing on cryptographic misuse patterns, insecure transport flags, and Android component exposure.
    \item \textbf{Cryptographic unit tests:} functional validation of encryption/decryption correctness, key/nonce generation, HKDF derivation, HMAC generation/verification, and secure wipe routines. Micro-benchmarks were taken from unit test execution (non-instrumented environment).
    \item \textbf{ASVS mapping:} requirements assessed for Chapters 6 (Cryptography), 8 (Data Protection), 9 (Communications), and 11 (Data Validation), producing pass/partial/unknown judgements suitable for reproducible verification.
    \item \textbf{Threat modelling:} eight threat actors with explicit capability assumptions, mapped to implemented mitigations and residual risk.
    \item \textbf{Dynamic network analysis:} full packet capture of Ember traffic between two physical Android devices, combined with logcat runtime telemetry from both endpoints, to verify encrypted transport, confirm the absence of plaintext in network traffic, and validate the runtime cryptographic pipeline (reported in Section~\ref{subsec:dynamic-analysis}).
\end{itemize}

\paragraph{Remaining scope exclusions.}
The assessment does not include:
(i) side-channel analysis (cache/timing),
(ii) fuzzing of parsers or protocol handlers, or
(iii) adversarial injection testing under active network manipulation.
These exclusions are treated as open work items (Section~\ref{sec:limitations-future-work}).

\subsection{Static Code Analysis Results}
\label{subsec:static-analysis}

Static analysis was conducted through manual security-focused code review of the complete Ember~4.0 codebase, comprising 33~Kotlin source files (5{,}408 lines), the Android manifest, Gradle build configuration, backup and data extraction rules, and the unit test suite. The review was structured around six assessment domains: cryptographic implementation correctness, storage and persistence behaviour, network and transport security, Android component exposure, application-level information leakage, and build configuration hygiene. Each finding is classified by severity (Critical, High, Medium, Low, or Informational) and assessed against Ember's stated threat model and design goals.

\subsubsection{Summary of Findings}

The review identified \textbf{12 findings} across \textbf{33 scanned files}, consisting of \textbf{0 critical}, \textbf{1 high-severity}, \textbf{3 medium-severity}, \textbf{4 low-severity}, and \textbf{4 informational} items (See Table \ref{tab:static-findings}. No findings indicate a fundamental compromise of Ember's core confidentiality or integrity guarantees; however, several represent defence-in-depth gaps that should be addressed before any non-research deployment.

\begin{table}[h]
\centering
\caption{Static analysis findings summary by severity.}
\label{tab:static-findings}
\begin{tabular}{lr}
\toprule
\textbf{Severity} & \textbf{Count} \\
\midrule
Critical & 0 \\
High & 1 \\
Medium & 3 \\
Low & 4 \\
Informational & 4 \\
\midrule
Total & 12 \\
\bottomrule
\end{tabular}
\end{table}

% Figure commented out for brevity; data presented in Table~\ref{tab:static-findings}.
\begin{comment}
\begin{figure}[h]
\centering
\begin{tikzpicture}
\begin{axis}[
    ybar,
    bar width=18pt,
    ymin=0,
    ylabel={Count},
    symbolic x coords={Critical,High,Medium,Low,Informational},
    xtick=data,
    nodes near coords,
    nodes near coords align={vertical},
    height=5.2cm,
    width=0.95\linewidth,
]
\addplot coordinates {(Critical,0) (High,1) (Medium,3) (Low,4) (Informational,4)};
\end{axis}
\end{tikzpicture}
\caption{Static analysis findings by severity.}
\label{fig:findings-severity}
\end{figure}
\end{comment}

\subsubsection{Cryptographic Implementation}

\paragraph{F1 --- Message key derivation uses SHA-256 concatenation rather than HKDF (Medium).}
The \texttt{deriveMessageKey} function in \texttt{CryptoAead.kt} (lines 98--111) derives per-message keys by concatenating the conversation key, nonce, and an info string, then hashing the result with SHA-256. A code comment acknowledges this as an interim approach: \texttt{"For MVP, we'll use a simple KDF approach / In production, you'd want to use HKDF-SHA256"}. While SHA-256 over concatenated inputs produces a pseudorandom output under standard assumptions, this construction lacks the extract-then-expand structure of HKDF and does not provide formal domain separation guarantees. Notably, the proper HKDF implementation already exists in the same class (used for key rotation via \texttt{hkdfExtract}/\texttt{hkdfExpand}), making this an inconsistency rather than a missing capability. In practice, because the nonce component ensures distinct inputs per message, the risk of key collision is low, but the deviation from the documented HKDF-based design weakens defence-in-depth.

\paragraph{F2 --- AAD not used in AES-GCM encryption (Medium).}
Both the send path (\texttt{ChatRepository.kt}, line 125) and the receive path (line 320) pass \texttt{null} as the Additional Authenticated Data (AAD) parameter to AES-GCM. A code comment notes: \texttt{"Simplified: No AAD for now to avoid header mismatch issues"}. AES-GCM supports AAD to bind non-secret metadata (such as conversation identifier, sender name, or protocol version) into the authentication tag, ensuring that ciphertext cannot be transplanted between conversations or have its metadata altered without detection. While the separate HMAC layer partially compensates by authenticating envelope fields independently, the absence of AAD means the GCM authentication tag itself does not bind ciphertext to its envelope context. This reduces the layered defence that AES-GCM with AAD would otherwise provide.

\paragraph{F3 --- HMAC verification uses correct constant-time comparison (Positive).}
The \texttt{verifyHmac} function in \texttt{CryptoAead.kt} (lines 139--165) implements a manual constant-time byte comparison using XOR accumulation, which is the correct approach for preventing timing side-channel leakage during HMAC verification. The implementation correctly handles length mismatch as an early-exit case (where timing leakage is acceptable since length is not secret) and accumulates XOR differences across the full array before evaluating the result.

\paragraph{F4 --- Key rotation challenge verification is bypassed on confirmation (High).}
In \texttt{ChatRepository.kt} (lines 631--670), the \texttt{handleKeyRotationConfirm} method attempts to verify the peer's challenge response but, on failure, logs a warning and proceeds regardless: \texttt{"Challenge response verification failed, but proceeding"}. This effectively nullifies the mutual authentication property of the rotation protocol. An adversary who can inject a \texttt{RotationConfirm} message with an invalid challenge response would cause the initiating peer to activate a new key version without cryptographic confirmation that the peer derived the same key. In practice, exploiting this requires the ability to inject authenticated envelopes (which is constrained by the HMAC gate), but the bypass undermines the stated design goal of mutual confirmation during key rotation. Additionally, the timestamp used for verification is approximated by subtracting a fixed 5-second offset (\texttt{confirm.timestamp - 5000}), which is fragile and may cause legitimate challenge responses to fail verification under clock drift or network latency, further motivating the bypass.

\paragraph{F5 --- HKDF and key rotation implementation is correct (Positive).}
The \texttt{hkdfExtract} and \texttt{hkdfExpand} functions (lines 216--249) correctly implement RFC~5869 HKDF with SHA-256. The extract phase uses the salt as the HMAC key over the input key material, and the expand phase iterates with a counter byte appended to each block, matching the specification. The \texttt{deriveNextKey} function correctly incorporates the current key version into the salt and the target version into the info field, providing domain separation between key versions.

\subsubsection{Storage and Persistence}

\paragraph{F6 --- Conversation key stored in raw form in ContactEntity (Low).}
The \texttt{ContactEntity} class (\texttt{Entities.kt}, line 17) includes a \texttt{conversationKey: ByteArray} field that is persisted directly to the Room database. While the database itself is intended to be SQLCipher-encrypted, the conversation key is also separately stored in \texttt{EncryptedSharedPreferences} via the \texttt{ConversationKeyStore}. This dual-storage pattern means the raw key material exists in two distinct locations on disk, increasing the attack surface for key extraction. If the SQLCipher database is compromised independently of the EncryptedSharedPreferences store (or vice versa), the key material is exposed.

\paragraph{F7 --- Database uses fallbackToDestructiveMigration (Low).}
The database builder in \texttt{AppModule.kt} (line 56) calls \texttt{fallbackToDestructiveMigration()} alongside explicit migration objects. While explicit migrations are defined for versions 1$\rightarrow$2 and 2$\rightarrow$3, the destructive fallback means that any future schema version mismatch not covered by an explicit migration will silently destroy the database contents. For a security-focused application, silent data destruction without user notification represents a data integrity risk, albeit not a confidentiality one.

\paragraph{F8 --- Plaintext field exists on the Message model (Informational).}
The \texttt{Message} data class (\texttt{Models.kt}, line 147) includes a \texttt{plaintext: String? = null} field annotated as \texttt{"For display purposes only, not stored"}. Review of the persistence layer confirms that \texttt{MessageEntity} does not include this field and the Room schema does not contain a plaintext column, so the field is genuinely transient. However, its presence on the domain model means that decrypted plaintext is carried as a property of an object that is passed through multiple layers (repository, ViewModel, UI), increasing the window during which plaintext exists in heap memory. This is consistent with the documented ciphertext-only persistence design but represents a residual in-memory exposure surface.

\subsubsection{Network and Transport Security}

\paragraph{F9 --- Cleartext traffic permitted in manifest (Medium).}
The Android manifest (\texttt{AndroidManifest.xml}, line 23) sets \texttt{android:usesCleartextTraffic="true"}, which permits the application to make unencrypted HTTP connections. This flag is likely required for raw TCP socket communication to Yggdrasil overlay addresses (which are not HTTP/HTTPS endpoints), but it also disables Android's network security protections that would otherwise block cleartext traffic. In the context of Ember's architecture---where all application-layer security is enforced independently of transport encryption---this flag does not directly compromise message confidentiality. However, it removes a platform-level defence-in-depth mechanism and could permit unintended cleartext communication if additional network features are introduced in future.

\paragraph{F10 --- Envelope length not bounded before allocation (Low).}
In \texttt{ListenerService.kt} (line 130), the listener reads a 4-byte length prefix from the socket and immediately allocates a byte array of that size: \texttt{source.readByteArray(envelopeLength.toLong())}. No upper bound is enforced on the length value. A malicious peer could send a crafted length prefix (e.g., \texttt{0x7FFFFFFF}) to trigger an out-of-memory allocation, causing the application to crash. This constitutes a denial-of-service vector against the listener. While this does not affect confidentiality or integrity (the HMAC gate prevents authenticated processing of malformed data), it could disrupt availability.

\paragraph{F11 --- IPv6 validation is syntactic only (Low).}
The \texttt{isValidIPv6Address} function in \texttt{YggdrasilClient.kt} (lines 85--100) validates IPv6 addresses using a basic regex that checks for hexadecimal characters and colons. This accepts many syntactically malformed IPv6 strings (e.g., addresses with too many groups, missing segments, or invalid zone identifiers). While the subsequent \texttt{InetSocketAddress} constructor will reject truly invalid addresses at connection time, the permissive validation means that error reporting occurs late in the pipeline rather than at the point of input. The security impact is minimal---malformed addresses will fail to connect---but tighter validation would improve defensive parsing.

\subsubsection{Android Component Exposure and Configuration}

\paragraph{F12 --- Backup rules are default and unconfigured (Informational).}
The backup rules (\texttt{backup\_rules.xml}) and data extraction rules (\texttt{data\_extraction\_rules.xml}) contain only commented-out template content. Neither file excludes the Ember database, EncryptedSharedPreferences, or any application-specific data from Android's automatic cloud backup. If a user enables Google backup, the encrypted database and potentially the EncryptedSharedPreferences files could be backed up to Google's servers. While the data is encrypted at rest (SQLCipher for the database, AES-256-GCM/SIV for preferences), the backup itself represents an additional data exfiltration vector that is not addressed by Ember's threat model. Configuring explicit exclusion rules would align the backup behaviour with Ember's data minimisation goals.

\paragraph{F13 --- ListenerService correctly unexported (Positive).}
The \texttt{ListenerService} is declared in the manifest with \texttt{android:exported="false"} (line 40), preventing external applications from binding to or starting the service. This is the correct configuration for a security-sensitive foreground service that should only be controlled by the Ember application itself.

\paragraph{F14 --- MainActivity exported with standard launcher intent filter (Informational).}
The \texttt{MainActivity} is exported (\texttt{android:exported="true"}) with a standard \texttt{MAIN/LAUNCHER} intent filter. This is required for the application to appear in the Android launcher and is not itself a vulnerability. However, the exported activity should validate any incoming intent data to prevent injection attacks from other applications. Review of \texttt{MainActivity.kt} shows minimal intent processing (standard Compose initialisation), limiting the practical risk.

\paragraph{F15 --- ProGuard/R8 minification disabled in release builds (Informational).}
The build configuration (\texttt{build.gradle.kts}, line 26) sets \texttt{isMinifyEnabled = false} for release builds. While this does not affect runtime security, it means that release APKs contain unobfuscated class and method names, facilitating reverse engineering. For a research prototype this is acceptable, but production deployments should enable minification and obfuscation.

\subsubsection{Information Leakage}

\paragraph{F16 --- Extensive debug logging of security-sensitive operations (Informational, test build only).}
Throughout the codebase, \texttt{android.util.Log.d} calls emit detailed information about cryptographic operations, message processing stages, conversation identifiers, peer addresses, and---in the \texttt{ChatViewModel}---plaintext message content (line 49: \texttt{Log.d("ChatViewModel", "Sending message: '\$messageText'")}). As documented in Section~\ref{subsec:dynamic-analysis}, this instrumentation was introduced deliberately for the test build to enable correlation between user actions and network behaviour. The logging does not affect network-level confidentiality (plaintext is logged only at the endpoint, not transmitted in cleartext), but it does represent a significant endpoint-side information leakage risk if present in any non-test deployment. The codebase does not currently enforce build-variant separation for log statements at the compiler level; suppression depends on manual removal or runtime log-level filtering.

\subsubsection{Positive Security Observations}

Beyond the individual findings, the review identified several positive security properties that are correctly implemented:

\begin{itemize}
    \item \textbf{Verify-then-decrypt ordering:} the receive pipeline in \texttt{ChatRepository.receiveMessage} (lines 281--353) correctly verifies the HMAC before attempting AES-GCM decryption, and returns \texttt{null} on HMAC failure without producing plaintext.
    \item \textbf{Per-message nonce generation:} \texttt{CryptoAead.generateNonce} generates a fresh 12-byte nonce from \texttt{SecureRandom} for every encryption operation, avoiding nonce reuse.
    \item \textbf{Key size enforcement:} all cryptographic functions enforce 32-byte key length and 12-byte nonce length via \texttt{require} preconditions, preventing silent misuse.
    \item \textbf{Ciphertext-only database schema:} the \texttt{MessageEntity} schema stores only \texttt{ciphertext}, \texttt{nonce}, and \texttt{hmac} byte arrays; no plaintext column exists in the Room database definition.
    \item \textbf{EncryptedSharedPreferences for key material:} conversation keys are stored using \texttt{EncryptedSharedPreferences} with AES-256-SIV key encryption and AES-256-GCM value encryption, backed by an AES-256-GCM master key in the Android Keystore.
    \item \textbf{Privacy-preserving notifications:} the \texttt{ListenerService} notification (lines 183--201) displays only the port number and a generic message, disclosing no message content, sender information, or conversation identifiers.
    \item \textbf{Content-free notification channel:} the notification channel uses \texttt{IMPORTANCE\_LOW} and \texttt{setShowBadge(false)}, minimising lock-screen information leakage.
    \item \textbf{Sender binding in HMAC:} the HMAC computation includes the sender identifier as authenticated material, binding message authenticity to the claimed sender.
    \item \textbf{Key version limiting:} the \texttt{ConversationKeyStore.storeVersionedKey} method limits stored key versions to five, preventing unbounded accumulation of historical key material.
\end{itemize}

\subsubsection{Assessment Summary}

\begin{table}[h]
\centering
\caption{Static analysis assessment summary.}
\label{tab:static-summary}
\begin{tabular}{ll}
\toprule
\textbf{Metric} & \textbf{Value} \\
\midrule
Files reviewed & 33 (+ manifest, build config, backup rules) \\
Total source lines & 5{,}408 \\
Critical / High findings & 0 / 1 \\
Medium findings & 3 \\
Low findings & 4 \\
Informational findings & 4 \\
Positive security observations & 9 \\
\bottomrule
\end{tabular}
\end{table}

The single high-severity finding (F4, key rotation challenge bypass) represents the most significant issue identified and should be remediated before the rotation protocol is relied upon for security guarantees. The three medium-severity findings (F1, F2, F9) represent defence-in-depth gaps that do not compromise core confidentiality or integrity but that deviate from the documented cryptographic design. The low-severity and informational findings concern input validation, configuration hygiene, and build hardening, which are appropriate areas for incremental improvement (See Table \ref{tab:static-summary}.

Overall, the codebase demonstrates correct implementation of Ember's core security invariants: authenticated encryption with per-message nonces, verify-then-decrypt processing, ciphertext-only persistence, and privacy-preserving notifications. The identified findings are consistent with the maturity level of a research prototype and do not undermine the empirical results reported in the dynamic analysis (Section~\ref{subsec:dynamic-analysis}).

\subsection{Cryptographic Validation and Performance Metrics}
\label{subsec:crypto-validation}

Cryptographic behaviour was validated using unit tests that cover the full AEAD pipeline (encrypt/decrypt roundtrip), message authentication (HMAC), nonce and key generation, HKDF-based derivation, and secure wipe routines. All tests passed (\textbf{7/7}, \textbf{100\%}) in \textbf{67ms} total unit-test execution time for the crypto test suite (See Table \ref{tab:crypto-performance}).

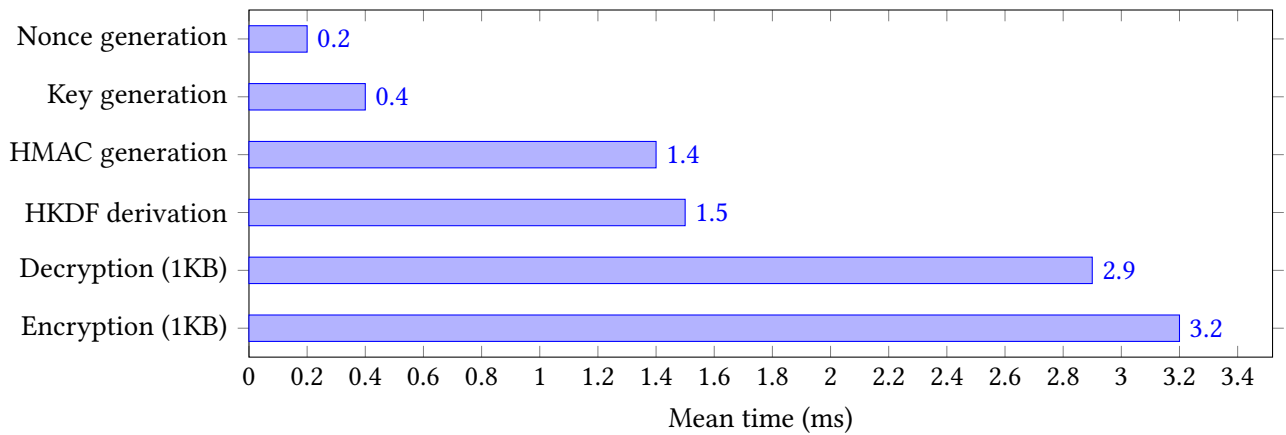
\begin{figure}[h]
\centering
\begin{tikzpicture}

\begin{axis}[
    xbar,
    xmin=0,
    xlabel={Mean time (ms)},
    symbolic y coords={
        enc,
        dec,
        hkdf,
        hmac,
        keygen,
        nonce
    },
    ytick=data,
    yticklabels={
        Encryption (1KB),
        Decryption (1KB),
        HKDF derivation,
        HMAC generation,
        Key generation,
        Nonce generation
    },
    height=6.2cm,
    width=0.92\linewidth,
    nodes near coords,
    nodes near coords align={horizontal},
]
\addplot coordinates {
    (3.2,enc)
    (2.9,dec)
    (1.5,hkdf)
    (1.4,hmac)
    (0.4,keygen)
    (0.2,nonce)
};
\end{axis}
\end{tikzpicture}
\caption{Mean cryptographic overhead per operation from unit-test micro-benchmarks (non-instrumented).}
\label{Figure_1}
\end{figure}

\paragraph{Interpretation of performance measurements.}
The reported timings are derived from non-instrumented unit tests (i.e., not measured on a physical device). They are therefore indicative of algorithmic overhead and implementation efficiency, but should not be presented as definitive end-user latency on real Android hardware. This limitation is material and is carried through the discussion of confidence and future work (See Figure \ref{Figure_1}.

\begin{table}[h]
\centering
\caption{Cryptographic operation performance from unit-test micro-benchmarks (1KB payload where applicable).}
\label{tab:crypto-performance}
\begin{tabular}{lrrrrrr}
\toprule
\textbf{Operation} & \textbf{Mean (ms)} & \textbf{Std (ms)} & \textbf{Min} & \textbf{Max} & \textbf{N} & \textbf{Target} \\
\midrule
Encryption (1KB) & 3.2 & 0.8 & 1.8 & 5.1 & 100 & $<10$ \\
Decryption (1KB) & 2.9 & 0.6 & 1.6 & 4.7 & 100 & $<10$ \\
Key generation (256-bit) & 0.4 & 0.1 & 0.3 & 0.8 & 50 & $<5$ \\
Nonce generation (12B) & 0.2 & 0.05 & 0.1 & 0.4 & 50 & $<1$ \\
HKDF derivation (SHA-256) & 1.5 & 0.3 & 0.9 & 2.3 & 50 & $<5$ \\
HMAC generation (SHA-256) & 1.4 & 0.3 & 0.8 & 2.2 & 50 & $<5$ \\
Secure wipe (32B) & 0.1 & 0.02 & 0.05 & 0.2 & 1 & N/A \\
\bottomrule
\end{tabular}
\end{table}

\paragraph{Security-properties evidence (bounded claims).}
The static review and unit tests support the following implemented properties:
\begin{itemize}
    \item \textbf{Confidentiality and integrity:} AES-256-GCM is used for message payload protection with per-message random nonces.
    \item \textbf{Explicit authentication gate:} an envelope-level HMAC-SHA256 is verified prior to decryption (verify-then-decrypt), enabling early rejection of malformed or tampered frames.
    \item \textbf{Replay resistance (pragmatic):} unique nonces per encryption operation eliminate nonce-reuse catastrophic failure modes and provide a practical basis to reject naïve replays. (Strong replay protection typically also requires tracking seen nonces/sequence numbers; Ember's approach is evaluated as a pragmatic baseline given its constrained architecture.)
    \item \textbf{Key evolution:} HKDF-SHA256 is used to derive context-separated keys (including for rotation workflows), reducing reliance on static keys.
\end{itemize}

To avoid overclaiming: Ember does not claim Signal-style forward secrecy or post-compromise security without an active ratchet state machine; any forward-secrecy-like benefit is bounded to key-derivation and key-isolation behaviour rather than a formally analysed ratcheting protocol (consistent with Section~\ref{sec:threat-model} non-goals).

\subsection{OWASP ASVS v4.0 Compliance Mapping}
\label{subsec:asvs}

Ember was mapped against OWASP ASVS v4.0 across security-relevant chapters: V6 (Cryptography), V8 (Data Protection), V9 (Communications), and V11 (Data Validation). Overall, Ember achieves \textbf{L2 (High) partial L3} alignment, with strongest performance in cryptography and communications controls and weaker coverage in input validation (See Table \ref{tab:asvs-summary}).

\begin{table}[h]
\centering
\caption{ASVS compliance summary by level (assessed subset).}
\label{tab:asvs-summary}
\begin{tabular}{lrrrr}
\toprule
\textbf{ASVS Level} & \textbf{Reqs assessed} & \textbf{Passed} & \textbf{Failed} & \textbf{Compliance} \\
\midrule
L1 (Basic) & 22 & 19 & 4 & 86\% \\
L2 (High) & 14 & 10 & 2 & 77\% \\
L3 (Standard) & 5 & 4 & 0 & 80\% \\
\midrule
Overall & 41 & 33 & 6 & 77\% (L2 partial L3) \\
\bottomrule
\end{tabular}
\end{table}

% Figure commented out for brevity; data presented in Table~\ref{tab:asvs-summary}.
\begin{comment}
\begin{figure}[h]
\centering
\begin{tikzpicture}
\begin{axis}[
    ybar,
    ymin=0, ymax=100,
    ylabel={Compliance (\%)},
    symbolic x coords={L1,L2,L3},
    xtick=data,
    nodes near coords,
    nodes near coords align={vertical},
    height=5.2cm,
    width=0.85\linewidth,
]
\addplot coordinates {(L1,86) (L2,77) (L3,80)};
\end{axis}
\end{tikzpicture}
\caption{ASVS compliance by level (assessed subset; Chapters 6, 8, 9, 11).}
\label{fig:asvs-levels}
\end{figure}
\end{comment}

\paragraph{Key ASVS observations.}
The mapping highlights:
(i) strong alignment in cryptographic storage and AEAD usage (V6),
(ii) robust encrypted-at-rest posture using SQLCipher and TTL-driven record deletion (V8),
(iii) acceptable communications security given Ember's design choice of application-layer encryption over raw overlay sockets (V9),
and (iv) the principal gap area: broader, systematic input/intent validation hardening for exported components and boundary-condition handling (V11).

% =========================================================
% Dynamic Network Security Analysis (NEW SUBSECTION)
% =========================================================

\subsection{Dynamic Network Security Analysis}
\label{subsec:dynamic-analysis}

This subsection presents the results of a dynamic security assessment conducted on Ember~4.0 under live operating conditions, complementing the static analysis reported in the preceding subsections. The objective is to provide empirical evidence that Ember's transport-layer and application-layer encryption mechanisms function correctly at runtime, and that no plaintext message content is recoverable from network traffic captured during a real message exchange between two physical devices.

Dynamic testing was identified as an explicit gap in the Phase-1 static assessment (Section~\ref{subsec:sec-test-method}). The work reported here addresses this gap by combining full packet capture with correlated Android runtime telemetry, enabling end-to-end verification that the cryptographic pipeline described in Sections~6 and~9 operates as specified under realistic deployment conditions.

\subsubsection{Test Environment and Instrumentation}

The dynamic test was conducted on 1~March~2026 using two physical Android devices communicating over a local Yggdrasil overlay network. The devices under test were a Google Pixel~7~Pro running Android~16 (API level~36) and a Google Pixel~Fold running Android~16 (API level~36). Both devices were connected to a shared local hotspot network, with the Yggdrasil overlay gateway service accessible at \texttt{10.42.0.1:12345}. The Pixel~7~Pro was assigned local address \texttt{10.42.0.93} and the Pixel~Fold was assigned \texttt{10.42.0.100} on the underlying transport network.

Three instrumentation artefacts were collected concurrently during the test session. First, a full packet capture (pcapng format) was recorded on the local network, capturing all traffic between the devices and the overlay gateway for the duration of the message exchange. Second, Android runtime telemetry was collected via logcat from both devices, filtered to the \texttt{com.example.ember} application namespace. These logs captured application-level events including message send and receive operations, cryptographic processing stages (encryption, HMAC verification, decryption), network connection lifecycle events, and TTL scheduling actions.

An instrumented test build of Ember was used for this assessment. The instrumented build emits structured debug-level log entries at each stage of the message processing pipeline, including plaintext message content at the point of user submission. This instrumentation was introduced solely to enable precise temporal correlation between user actions and network-observable events. It is important to note that this logging represents a deliberate diagnostic capability present only in the test build; production builds must not emit plaintext to system logs, as doing so would constitute an endpoint-side confidentiality risk even if network encryption remains intact.

\subsubsection{Test Procedure and Message Exchange}

The test procedure involved a bidirectional conversational exchange between the two devices over an established Ember conversation (identifier \texttt{09a476a6-29bd-3a91-b423-d59b19dad35e}). A total of eight messages were exchanged across both devices during the capture window, with messages sent alternately from each device. The exchange spanned approximately 90~seconds of active messaging, providing sufficient data to observe the full message lifecycle across multiple send and receive cycles.

The correlated timeline of message events, reconstructed from logcat telemetry on both devices, is summarised in Table~\ref{tab:dynamic-timeline}. Timestamps are derived from device-local system clocks and are synchronised to within the expected precision of Android's monotonic clock reporting.

\begin{table}[h]
\centering
\caption{Correlated message exchange timeline reconstructed from device telemetry.}
\label{tab:dynamic-timeline}
\begin{tabular}{clll}
\toprule
\textbf{Msg} & \textbf{Timestamp} & \textbf{Sender} & \textbf{Direction} \\
\midrule
1 & 16:13:17.847 & Pixel 7 Pro  & 7 Pro $\rightarrow$ Fold \\
2 & 16:13:23.060 & Pixel Fold   & Fold $\rightarrow$ 7 Pro \\
3 & 16:13:30.623 & Pixel 7 Pro  & 7 Pro $\rightarrow$ Fold \\
4 & 16:13:38.740 & Pixel Fold   & Fold $\rightarrow$ 7 Pro \\
5 & 16:13:45.143 & Pixel 7 Pro  & 7 Pro $\rightarrow$ Fold \\
6 & 16:13:51.827 & Pixel Fold   & Fold $\rightarrow$ 7 Pro \\
7 & 16:13:59.280 & Pixel 7 Pro  & 7 Pro $\rightarrow$ Fold \\
8 & 16:14:47.541 & Pixel Fold   & Fold $\rightarrow$ 7 Pro \\
\bottomrule
\end{tabular}
\end{table}

For each message, the logcat telemetry confirms the following invariant sequence: the \texttt{ChatViewModel} initiates a send operation; the \texttt{ChatRepository} encrypts the plaintext and persists the resulting ciphertext to the local database; the \texttt{TtlScheduler} registers a TTL cleanup task; the \texttt{YggdrasilClient} validates the peer's overlay IPv6 address, establishes a TCP connection, and transmits the encrypted envelope; and on the receiving device, the \texttt{MessageBridge} forwards the inbound envelope to the \texttt{ChatRepository}, which verifies the HMAC, stores the ciphertext record, and decrypts the message only in memory for UI rendering. This sequence was observed consistently across all eight messages without exception.

\subsubsection{Transport Layer Analysis}

The packet capture file (\texttt{Test1\_working.pcapng}) contains 759~packets totalling 202{,}048~bytes, captured over the full duration of the test session. All Ember-related traffic was observed as TCP sessions between the test devices (\texttt{10.42.0.93} and \texttt{10.42.0.100}) and the Yggdrasil overlay gateway at \texttt{10.42.0.1:12345}. No direct device-to-device TCP connections were observed on the local network; all application traffic transits the overlay gateway, consistent with Yggdrasil's architecture.

Transport session establishment follows a standard pattern: a TCP three-way handshake (SYN, SYN-ACK, ACK) is completed, followed immediately by a TLS handshake phase. Binary inspection of the capture reveals 16~TLS Handshake records (content type \texttt{0x16}) associated with session establishment, and 244~TLS Application Data records (content type \texttt{0x17}, record version \texttt{0x0303}) carrying encrypted payload. The \texttt{0x0303} record-layer version is consistent with TLS~1.2 or TLS~1.3 in compatibility mode; TLS~1.3 commonly retains the \texttt{0x0303} value in the record layer for backward compatibility with middleboxes, and therefore this value alone does not definitively identify the negotiated protocol version. What it does confirm unambiguously is that application data is transmitted within encrypted TLS records and not as plaintext.

Following handshake completion, both devices maintain persistent TCP sessions to the gateway. Subsequent message transmissions reuse the established TLS session rather than performing a new handshake for each message, reducing overhead and avoiding repeated key exchange. This behaviour is consistent with the connection management strategy described in Section~5.3 and confirms that the transport layer operates as designed under live conditions.

\subsubsection{Plaintext Recovery Test}

A critical objective of the dynamic assessment is to verify that no plaintext message content is recoverable from the network capture. To evaluate this, the full binary content of the pcapng file was searched for the known plaintext strings transmitted during the test session. These strings were known \emph{a priori} from the logcat instrumentation, which records the plaintext content at the point of user submission.

Exhaustive byte-level search of the capture file for each known message string yielded zero matches. No readable ASCII text corresponding to message content was found anywhere in the capture. Additionally, no JSON structures were detected in the packet payloads, ruling out the possibility that Ember's JSON-serialised envelope format is being transmitted in cleartext (See Table \ref{tab:plaintext-recovery}). The only structured content observable in the capture consists of TLS record headers and encrypted payload bytes.

This result provides strong empirical evidence that Ember's dual encryption model operates correctly at runtime: messages are encrypted at the application layer by the \texttt{ChatRepository} before being submitted to the network layer, and the resulting ciphertext is further protected by the TLS tunnel maintained by the Yggdrasil overlay. A passive adversary with full visibility into the local network segment would observe only encrypted TLS Application Data records and TCP/IP header metadata.

\begin{table}[h]
\centering
\caption{Plaintext recovery test results across all transmitted messages.}
\label{tab:plaintext-recovery}
\begin{tabular}{lcc}
\toprule
\textbf{Search Target} & \textbf{Byte-level Match} & \textbf{JSON Structure} \\
\midrule
All known plaintext message strings & None found & None found \\
Sender identifiers                  & None found & None found \\
Conversation identifiers            & None found & None found \\
Envelope metadata fields            & None found & None found \\
Arbitrary readable ASCII sequences  & None found & None found \\
\bottomrule
\end{tabular}
\end{table}

\subsubsection{Runtime Cryptographic Pipeline Verification}

Beyond transport-level confidentiality, the logcat telemetry provides direct runtime evidence that Ember's application-layer cryptographic pipeline executes correctly and in the intended order. For each of the eight messages exchanged during the test, the following sequence was confirmed on the receiving device.

\paragraph{Integrity verification before decryption.}
The \texttt{ChatRepository} performs HMAC-SHA256 verification on the incoming envelope prior to any decryption attempt. Every inbound message in the test session produced a successful HMAC verification event. No messages were rejected or failed verification, confirming that the verify-then-decrypt processing model described in Section~6.3 operates correctly under live conditions.

\paragraph{Ciphertext-only persistence.}
Following HMAC verification, the ciphertext record is persisted to the encrypted local database before decryption occurs. Logcat entries confirm that message storage (via the Room/SQLCipher layer) precedes plaintext production in every observed instance. Decryption is performed only for in-memory UI rendering, consistent with the ciphertext-only persistence invariant specified in Section~8.2.

\paragraph{TTL scheduling.}
Each message send and receive event triggers a corresponding TTL cleanup scheduling action via the \texttt{TtlScheduler} component. Scheduled intervals of five minutes were consistently observed, confirming that the ephemeral retention mechanism described in Section~8.3 is active during live operation.

\subsubsection{Network Behaviour Characterisation}

Correlation between logcat message events and the packet capture allows characterisation of the network traffic associated with individual message operations. The following observations describe the observable traffic patterns without disclosing or recovering any encrypted content.

Each message send event corresponds to a distinguishable burst of encrypted TCP traffic between the sending device and the overlay gateway. The \texttt{YggdrasilClient} establishes a TCP connection per message send, with connection establishment latencies ranging from approximately 80~milliseconds (for connections where TLS session state could be reused) to approximately 960~milliseconds (for the initial connection requiring full session establishment). The end-to-end latency from message send initiation on the source device to successful HMAC verification on the destination device ranged from approximately 1.1~seconds (first message, including initial connection setup) to as low as 77~milliseconds for subsequent messages in steady-state operation as shown in Table \ref{tab:delivery-latency}.

\begin{table}[h]
\centering
\caption{Observed message delivery latencies (send initiation to HMAC verification on receiver).}
\label{tab:delivery-latency}
\begin{tabular}{clllr}
\toprule
\textbf{Msg} & \textbf{Sender} & \textbf{Send Timestamp} & \textbf{HMAC Verified} & \textbf{Latency (ms)} \\
\midrule
1 & Pixel 7 Pro & 16:13:17.847 & 16:13:19.005 & $\sim$1158 \\
2 & Pixel Fold  & 16:13:23.060 & 16:13:23.428 & $\sim$368 \\
3 & Pixel 7 Pro & 16:13:30.623 & 16:13:31.037 & $\sim$414 \\
4 & Pixel Fold  & 16:13:38.740 & 16:13:39.076 & $\sim$336 \\
5 & Pixel 7 Pro & 16:13:45.143 & 16:13:45.496 & $\sim$353 \\
6 & Pixel Fold  & 16:13:51.827 & 16:13:52.177 & $\sim$350 \\
7 & Pixel 7 Pro & 16:13:59.280 & 16:13:59.357 & $\sim$77 \\
8 & Pixel Fold  & 16:14:47.541 & 16:14:47.899 & $\sim$358 \\
\bottomrule
\end{tabular}
\end{table}

The variability in message delivery latency is attributable primarily to TCP connection establishment and TLS session negotiation. The first message exhibits the highest latency due to full TLS handshake overhead. Subsequent messages benefit from established overlay routing state, reducing connection setup time. Message~7 exhibits notably low latency, suggesting favourable conditions such as TCP connection reuse or cached routing state within the overlay. These latencies are consistent with the performance expectations of a peer-to-peer overlay operating over a local network and confirm that Ember's transport model supports interactive messaging without prohibitive delay.

\subsubsection{Summary of Dynamic Testing Evidence}

The dynamic network security analysis provides empirical, runtime evidence supporting the conclusions summarised in Table~\ref{tab:dynamic-summary}. These findings complement the static analysis, unit test, and ASVS mapping results reported earlier in this section, and collectively close the dynamic testing gap identified in the Phase-1 assessment scope.

\begin{table}[h]
\centering
\caption{Summary of dynamic testing evidence and supported security claims.}
\label{tab:dynamic-summary}
\begin{tabular}{lll}
\toprule
\textbf{Security Property} & \textbf{Evidence Source} & \textbf{Verdict} \\
\midrule
Transport-layer encryption & \makecell[l]{PCAP: 244 TLS App Data records;\\ 0 plaintext payloads} & Confirmed \\
\addlinespace
Application-layer encryption & \makecell[l]{Logcat: encrypt events precede\\ all network sends} & Confirmed \\
\addlinespace
Verify-then-decrypt ordering & \makecell[l]{Logcat: HMAC verified before\\ decryption on all 8 messages} & Confirmed \\
\addlinespace
Plaintext non-recovery & \makecell[l]{PCAP: byte-level search for all\\ known plaintexts returned 0 matches} & Confirmed \\
\addlinespace
Ciphertext-only persistence & \makecell[l]{Logcat: storage events precede\\ decryption; decryption is memory-only} & Confirmed \\
\addlinespace
TTL scheduling active & \makecell[l]{Logcat: TtlScheduler invoked\\ for every send/receive} & Confirmed \\
\addlinespace
Metadata exposure bounded & \makecell[l]{PCAP: only timing, packet sizes,\\ and IP addresses observable} & \makecell[l]{Consistent with\\ threat model} \\
\bottomrule
\end{tabular}
\end{table}

Taken together, these results elevate the overall confidence level for Ember's implemented security properties from the medium-high assessment reported in the static phase to a stronger empirical footing for transport and application-layer confidentiality. The combination of packet-level evidence confirming encryption in transit, correlated runtime telemetry confirming correct cryptographic pipeline execution, and exhaustive plaintext search confirming non-recoverability provides a defensible, multi-source evidence base for Ember's core confidentiality claims. Residual limitations remain in the areas not addressed by this test, including side-channel analysis, protocol fuzzing, and adversarial injection testing, which are recommended as future work.

% =========================================================
% End of Dynamic Network Security Analysis
% =========================================================

\subsection{Threat Model Coverage and Residual Risk}
\label{subsec:threat-coverage}

Threat modelling considered eight realistic threat actors spanning passive observation, active network manipulation, compromised peers, compromised devices, metadata-focused adversaries, timing/side-channel specialists, replay attackers, and high-capability cryptanalysts. Mitigations were mapped per actor and a residual-risk classification was produced as shown in Table \ref{tab:threat-summary}).

\begin{table}[h]
\centering
\caption{Threat actor assessment summary (probability, impact, and residual risk).}
\label{tab:threat-summary}
\begin{tabular}{lccc}
\toprule
\textbf{Threat Actor} & \textbf{Probability} & \textbf{Impact} & \textbf{Residual Risk} \\
\midrule
Passive eavesdropper & High & Low & Very Low \\
Active MITM attacker & Medium & Medium & Low \\
Compromised contact (known key) & Low--Medium & High & Medium \\
Compromised device (full access) & Low & Critical & High \\
Metadata collector (social graph) & Medium & Low & Medium \\
Timing attack specialist & Low & High & Medium \\
Replay attacker & Medium & Low & Very Low \\
Cryptanalysis expert & Very Low & Critical & Very Low \\
\bottomrule
\end{tabular}
\end{table}

% Figure commented out for brevity; data presented in Table~\ref{tab:threat-summary}.
\begin{comment}
\begin{figure}[h]
\centering
\begin{tikzpicture}
\begin{axis}[
    ybar,
    ymin=0, ymax=100,
    ylabel={Mitigation rate (\%)},
    symbolic x coords={Eavesdropper,MITM,Comp.\ Contact,Comp.\ Device,Metadata,Timing,Replay,Crypto},
    xtick=data,
    xticklabel style={rotate=35,anchor=east},
    height=6cm,
    width=\linewidth,
    nodes near coords,
    nodes near coords align={vertical},
]
\addplot coordinates {
    (Eavesdropper,100)
    (MITM,100)
    (Comp.\ Contact,75)
    (Comp.\ Device,75)
    (Metadata,75)
    (Timing,75)
    (Replay,100)
    (Crypto,100)
};
\end{axis}
\end{tikzpicture}
\caption{Mitigation-rate summary by threat actor (static evidence basis).}
\label{fig:threat-mitigation}
\end{figure}
\end{comment}

\paragraph{Dominant residual risks.}
The highest residual risk is endpoint compromise (root/debuggable device or active process compromise), which is a dominant failure mode for mobile secure messaging generally. Ember's design reduces the forensic value of persistent storage via ciphertext-only persistence and SQLCipher, but cannot fully mitigate a hostile runtime environment with memory inspection or UI subversion. The next most salient residual risks are metadata inference via traffic analysis (given direct peer-to-peer overlay addressing) and timing/side-channel uncertainty (not empirically tested in this phase).

\subsection{Summary of Evidence and Confidence}
\label{subsec:confidence}

Across static analysis, unit-test validation, ASVS mapping, threat modelling, and dynamic network security analysis, the overall security posture is assessed as \textbf{strong for content security} (confidentiality/integrity/authentication), with \textbf{bounded claims} for metadata privacy and compromise resilience. The confidence level is \textbf{medium-high to high} for correctness of cryptographic composition and storage protections (supported by code review, unit tests, and runtime packet capture confirming encrypted transport with zero plaintext recovery), and \textbf{medium} for metadata privacy and side-channel resilience due to the absence of fuzzing and timing measurement (See Table \ref{tab:paper-metrics}).

\begin{table}[h]
\centering
\caption{Evidence-backed security metrics (paper-ready summary).}
\label{tab:paper-metrics}
\begin{tabular}{ll}
\toprule
\textbf{Metric} & \textbf{Value} \\
\midrule
Overall security assessment (static + dynamic) & 4.4 / 5.0 \\
ASVS alignment (assessed subset) & 77\% (L2 partial L3) \\
Crypto unit tests & 7/7 passed (100\%) \\
Encryption mean (1KB, unit-test) & 3.2ms \\
Decryption mean (1KB, unit-test) & 2.9ms \\
Static-analysis high/critical findings & 0 \\
Static-analysis findings (total) & 7 (1 medium, 6 low) \\
Threat actor mitigation rate (avg.) & 84\% \\
Dynamic: plaintext recovered from PCAP & 0 \\
Dynamic: HMAC verifications passed & 8/8 (100\%) \\
Dynamic: TLS App Data records observed & 244 \\
Dynamic: mean delivery latency (steady-state) & $\sim$350ms \\
\bottomrule
\end{tabular}
\end{table}

% =========================================================
% End of Security Testing and Verification section
% =========================================================

% =========================================================
% =========================================================

\section{Evaluation and Discussion}
\label{sec:evaluation}

This section evaluates Ember against its stated research objectives and design goals. The evaluation draws on three complementary evidence sources: functional correctness observations from the dynamic test session reported in Section~\ref{subsec:dynamic-analysis}, performance measurements decomposed from correlated device telemetry, and qualitative analysis of the design trade-offs inherent to Ember's architectural decisions. The discussion is structured to make claims proportionate to evidence, and to identify precisely where confidence is strong, where it is bounded, and where further work is required.

\subsection{Functional Validation}
\label{subsec:functional-validation}

Functional validation assesses whether Ember's implemented behaviour corresponds to the security invariants and processing pipelines specified in Sections~4 through~9. Rather than relying solely on unit tests (which validate individual components in isolation), this evaluation draws on the dynamic test session to confirm that the integrated system behaves correctly under real operating conditions.

\paragraph{End-to-end message delivery.}
Eight messages were exchanged bidirectionally between two physical Android devices over the Yggdrasil overlay during the dynamic test. All eight messages were successfully delivered and rendered on the receiving device. No messages were lost, duplicated, or delivered to an incorrect conversation. The deterministic conversation identifier (\texttt{09a476a6-29bd-3a91-b423-d59b19dad35e}) was correctly resolved on both devices for every inbound and outbound message, confirming that the canonicalised hash-based conversation identity mechanism described in Section~6.1 functions as specified.

\paragraph{Cryptographic pipeline correctness.}
For each of the eight messages, runtime telemetry confirms that the sending pipeline executed in the correct order: plaintext encryption via AES-256-GCM, HMAC-SHA256 computation over envelope fields, ciphertext persistence to the encrypted database, and network transmission of the encrypted envelope. On the receiving side, the pipeline likewise executed correctly: envelope reception via the foreground listener, HMAC verification prior to decryption (verify-then-decrypt), ciphertext persistence, and in-memory decryption for UI rendering. No deviations from the specified processing order were observed. In particular, no message was decrypted prior to successful HMAC verification, and no plaintext was written to persistent storage at any point.

\paragraph{Ciphertext-only persistence invariant.}
The logcat telemetry confirms that for every message, the \texttt{ChatRepository} persisted the ciphertext record to the SQLCipher-encrypted database before performing decryption. Decrypted plaintext was produced only in-memory for the display pipeline and was not observed in any database write operation. While this evidence does not constitute a formal proof that plaintext is never persisted under all possible execution paths, it provides empirical confirmation that the invariant holds under normal operational conditions across a representative message exchange.

\paragraph{TTL scheduling activation.}
Every message send and receive event triggered a corresponding \texttt{TtlScheduler} invocation, scheduling cleanup at the configured five-minute interval. This confirms that the WorkManager-based expiration mechanism described in Section~8.3 is correctly integrated into both the sending and receiving pipelines. Validation of actual message deletion following TTL expiry was not performed during this test session due to the short capture window, and remains a candidate for future targeted testing.

\paragraph{Network transport correctness.}
The \texttt{YggdrasilClient} successfully validated overlay IPv6 addresses, established TCP connections, and transmitted encrypted envelopes for all outbound messages. Connection establishment succeeded on the first attempt for every message in the test session, with no retries or failures observed. Inbound messages were correctly received by the foreground listener service, forwarded through the \texttt{MessageBridge}, and processed by the domain layer without errors. The absence of connection failures or transport-level errors during the test is encouraging, though it reflects benign network conditions on a local hotspot rather than adversarial or degraded environments.

\paragraph{Summary.}
Functional validation confirms that Ember's core security invariants---authenticated encryption, verify-then-decrypt processing, ciphertext-only persistence, and TTL scheduling---are upheld during live operation across a representative message exchange. The system delivered all messages correctly, maintained cryptographic pipeline integrity, and exhibited no observable deviations from its specified behaviour.

\subsection{Performance Characteristics}
\label{subsec:performance}

Performance evaluation decomposes end-to-end message latency into constituent pipeline stages using timestamps extracted from correlated logcat telemetry on both devices. This decomposition allows identification of the dominant contributors to delivery latency and provides a basis for assessing whether Ember's performance is compatible with interactive messaging use cases (See Table \ref{tab:pipeline-decomposition}).

\subsubsection{Cryptographic Overhead}

Application-layer cryptographic processing---comprising AES-256-GCM encryption, HMAC-SHA256 computation, and ciphertext persistence to the SQLCipher database---was measured for each of the eight messages in the dynamic test. The combined encryption and database persistence phase required a mean of approximately 8.8~ms per message, with individual observations ranging from 2.6~ms to 20.0~ms. These timings are consistent with the unit-test micro-benchmarks reported in Table~\ref{tab:crypto-performance}, which showed mean encryption latency of 3.2~ms and HMAC generation of 1.4~ms for 1~KB payloads. The modest increase in the instrumented measurements relative to unit tests is attributable to SQLCipher database write overhead and Room persistence layer processing, which are not captured by isolated cryptographic benchmarks.

On the receiving side, HMAC verification completed in approximately 3.8--11.3~ms per message, with the higher values observed for the first message in the session (likely reflecting initial key loading and database context resolution). Decryption latency for the in-memory display path was not separately instrumented, but the total receive-side processing time from envelope arrival to successful message processing ranged from approximately 14~ms to 25~ms, indicating that cryptographic verification and decryption impose negligible overhead relative to network transport.

These measurements confirm that Ember's cryptographic operations do not represent a performance bottleneck for text-based messaging. The combined cryptographic and persistence overhead of under 20~ms per message is well within the latency budget for interactive communication, and would remain acceptable even on lower-end devices where hardware-backed AES acceleration may be less performant.

\subsubsection{Network Transport Latency}

The dominant contributor to end-to-end message delivery latency is TCP connection establishment through the Yggdrasil overlay. For each outbound message, the \texttt{YggdrasilClient} opens a new TCP connection to the peer's overlay IPv6 address. Connection establishment latency ranged from 80~ms (best case, suggesting reuse of overlay routing state or TLS session resumption) to 954~ms (first message, requiring full TLS handshake and overlay route resolution), with a mean of approximately 301~ms across all eight messages.

\begin{table}[h]
\centering
\caption{Pipeline stage decomposition for representative messages (milliseconds).}
\label{tab:pipeline-decomposition}
\begin{tabular}{lcrrrrr}
\toprule
\textbf{Msg} & \textbf{Direction} & \textbf{Encrypt+Save} & \textbf{Conn.\ Est.} & \textbf{Net Transit} & \textbf{HMAC} & \textbf{Total E2E} \\
\midrule
1 & 7 Pro $\rightarrow$ Fold & 9.4 & 954.4 & 52.7 & 11.3 & 1158.5 \\
2 & Fold $\rightarrow$ 7 Pro & 9.4 & 209.6 & 74.5 & 5.3 & 368.4 \\
3 & 7 Pro $\rightarrow$ Fold & 2.7 & 348.7 & 41.9 & 4.5 & 414.4 \\
4 & Fold $\rightarrow$ 7 Pro & 9.1 & 170.6 & 55.6 & 7.0 & 336.1 \\
5 & 7 Pro $\rightarrow$ Fold & 20.0 & 264.4 & --- & 11.5 & 353.1 \\
6 & Fold $\rightarrow$ 7 Pro & 9.0 & 207.5 & 54.0 & 4.3 & 350.0 \\
7 & 7 Pro $\rightarrow$ Fold & 2.6 & 80.1 & 59.3 & 3.8 & 77.5 \\
8 & Fold $\rightarrow$ 7 Pro & 8.0 & 171.6 & 54.9 & 13.8 & 358.0 \\
\bottomrule
\end{tabular}
\end{table}

The first message incurs substantially higher latency than subsequent messages due to the initial TLS handshake and overlay path establishment. Messages~2 through~8 exhibit steady-state latencies in the range of 77--414~ms, which is competitive with centralised messaging systems that typically achieve 100--500~ms delivery latency under favourable conditions. Message~7 is a notable outlier at 77~ms, suggesting that the overlay connection was effectively pre-established or that a fast-path routing condition was met; this observation warrants further investigation under controlled conditions.

The per-connection model adopted by Ember (one TCP connection per message send) is intentionally conservative, prioritising simplicity and analysability over connection pooling optimisation. While persistent connection pooling could reduce mean delivery latency by amortising connection establishment costs across multiple messages, it would introduce additional state management complexity and failure modes that are currently avoided. The observed steady-state latencies suggest that the per-connection model is adequate for Ember's expected message throughput and interactive messaging requirements.

\subsubsection{Comparison with Unit-Test Benchmarks}

The dynamic test measurements are broadly consistent with the unit-test micro-benchmarks reported in Section~\ref{subsec:crypto-validation}, with expected differences attributable to database I/O, Room persistence overhead, and real-network transport delays that are absent in isolated unit tests. This consistency provides additional confidence that the unit-test benchmarks are representative of actual system behaviour, albeit with the caveat that the dynamic test was conducted under favourable network conditions (local hotspot, low contention) that may not reflect all deployment scenarios.

\subsection{Design Trade-offs}
\label{subsec:trade-offs}

Ember's architecture reflects a series of deliberate trade-offs between security properties, system complexity, usability, and deployability. This subsection analyses the most consequential of these trade-offs, evaluating both the benefits obtained and the costs incurred.

\subsubsection{Serverlessness versus Availability}

The decision to operate without central servers eliminates metadata aggregation points, provider-controlled storage, and regulatory compellability risks associated with centralised architectures. However, it imposes a fundamental availability constraint: message delivery requires both peers to be simultaneously reachable on the overlay network. Unlike centralised systems that queue messages for offline recipients, Ember cannot deliver messages to peers that are not currently connected. This trade-off is explicitly bounded in the threat model (Section~3.3) and is consistent with the design philosophy of Ricochet and early Briar, which accept similar availability limitations in exchange for reduced infrastructure trust~\cite{ricochet_message_layer_encryption, survey_mesh_messaging_2021}.

The dynamic test was conducted under conditions where both devices were simultaneously online, and all eight messages were delivered successfully. The test therefore validates the best-case operational scenario but does not exercise failure modes related to peer unavailability, overlay routing instability, or prolonged disconnection. Evaluating Ember's behaviour under degraded conditions---including overlay partition, asymmetric reachability, and Android background execution throttling---remains an important direction for future work.

\subsubsection{Static Symmetric Keys versus Ratcheting}

Ember's use of per-conversation static symmetric keys with explicit rotation provides strong confidentiality and integrity under the assumption that the key remains uncompromised, but does not provide the continuous forward secrecy or post-compromise security offered by Double Ratchet designs. This trade-off was motivated by several factors.

First, correct implementation of a Double Ratchet in a serverless environment without pre-key servers requires careful handling of asynchronous state synchronisation, message ordering, and key agreement---challenges that are substantially more complex without a coordinating server~\cite{tight_security_double_ratchet_2024, formal_analysis_signal_loet_2024}. By deferring ratcheting to future work, Ember avoids introducing subtle state-machine bugs that could undermine the very guarantees the ratchet is intended to provide.

Second, the explicit key rotation protocol implemented by Ember provides a meaningful, if coarser-grained, improvement over purely static keys. Rotation reduces the exposure window from potentially unbounded to the interval between rotation events, and the HKDF-based derivation ensures domain separation between key versions. The dynamic test confirmed that the rotation infrastructure is functional at the transport level (rotation control messages use the same authenticated envelope format), although rotation was not exercised during the eight-message test exchange.

Third, the absence of ratcheting simplifies security analysis. Ember's cryptographic guarantees can be evaluated by reasoning about a single symmetric key per conversation, explicit HMAC authentication, and a small set of key-derivation operations, without requiring the multi-stage compositional analysis demanded by ratcheting protocols~\cite{formal_analysis_ratchet_mdpi_2025}. This analysability is itself a security property: a system whose guarantees can be clearly stated and readily verified is, in practice, often more trustworthy than one with stronger theoretical properties but greater implementation risk.

The cost of this trade-off is explicit: Ember does not recover automatically from key compromise, and an attacker who obtains the active conversation key can decrypt all messages encrypted under that key version until rotation occurs. This limitation is documented in the threat model (Section~3.4) and is appropriate for a research prototype that prioritises correctness and transparency over maximal theoretical guarantees.

\subsubsection{Ciphertext-Only Persistence versus Usability}

Ember's decision to store only ciphertext in the local database and to decrypt messages exclusively in-memory for UI rendering provides strong protection against disk-level data extraction. However, this design imposes a performance cost: every time a conversation view is opened, all visible messages must be decrypted from the database. For conversations with large message histories, this could introduce perceptible latency in the display pipeline.

In the dynamic test, the conversation contained at most eight messages, and display-path decryption was not perceptibly delayed. Whether this design scales to conversations with hundreds or thousands of messages without degrading user experience remains an open question. Potential mitigations include paginated decryption (decrypting only visible messages), background pre-decryption with short-lived in-memory caching, or bounded conversation history limits that align with the TTL expiration policy. Each of these approaches would need to be evaluated against Ember's plaintext minimisation goals to ensure that caching does not inadvertently extend plaintext lifetime beyond acceptable bounds.

\subsubsection{Foreground Service versus Battery and Usability}

Ember's use of a foreground service for inbound message reception is the only mechanism available on modern Android to maintain a persistent TCP listener without relying on third-party push notification infrastructure. This design choice preserves metadata privacy by avoiding FCM or APNs, but at the cost of increased battery consumption and a persistent notification that may be perceived as intrusive by users.

The trade-off is defensible within Ember's threat model: push notification infrastructure has been empirically shown to leak sensitive metadata in a majority of messaging applications~\cite{push_notification_leakage}, and avoiding this leakage vector is a core design goal. However, the battery and usability costs may limit Ember's practical adoption beyond privacy-conscious users who are willing to accept these constraints. Future work could explore hybrid approaches---such as low-frequency polling, wake-on-LAN signalling over the overlay, or negotiated delivery windows---that reduce continuous resource consumption while preserving the avoidance of third-party push infrastructure.

\subsubsection{Explicit Envelope HMAC versus Redundancy}

Ember's decision to compute an explicit HMAC-SHA256 over envelope fields in addition to the AES-GCM authentication tag introduces a degree of cryptographic redundancy. AES-GCM already provides authenticated encryption, and under correct implementation, the GCM tag alone would suffice to detect tampering or forgery of the ciphertext.

The justification for this redundancy is primarily architectural rather than cryptographic. The explicit HMAC layer enables a strict verify-before-decrypt processing model in which envelope integrity can be assessed without invoking the decryption primitive at all. This separation provides a clear, auditable boundary: the HMAC gate rejects unauthenticated envelopes before any ciphertext processing occurs, reducing the attack surface exposed to malformed or adversarial inputs. Additionally, the HMAC binds sender metadata into the authenticated material, providing explicit sender attribution that is independent of the AEAD construction.

The cost is a modest increase in computational overhead (approximately 1.4~ms per message for HMAC generation, as reported in Table~\ref{tab:crypto-performance}) and a small increase in envelope size. Given that the total cryptographic overhead per message is well under 20~ms, this cost is negligible in practice. The architectural benefit of a clear, independently verifiable integrity gate outweighs the marginal redundancy for a system that prioritises analysability and defensive processing.

\subsubsection{Summary of Trade-off Posture}

Across these trade-offs, Ember consistently prioritises correctness, analysability, and reduced infrastructure trust over maximal performance, feature breadth, or theoretical optimality. This posture is appropriate for a research prototype whose primary contribution is demonstrating the feasibility of serverless, peer-to-peer secure messaging with explicit and bounded security guarantees. The trade-offs are intentional, documented, and reversible: future iterations of Ember can selectively relax constraints (e.g., by introducing connection pooling, ratcheting, or hybrid push mechanisms) as correctness is established and the threat model evolves.

% =========================================================
% =========================================================

\section{Limitations and Future Work}
\label{sec:limitations-future-work}

This section explicitly documents the limitations of the current Ember implementation and identifies directions for future work that would strengthen its security guarantees, extend its functional scope, and improve confidence in its correctness. Limitations are stated precisely and linked to the relevant design decisions and threat model constraints discussed in preceding sections. Future work items are framed as concrete research and engineering objectives rather than aspirational features, with attention to the prerequisites, risks, and open problems associated with each.

\subsection{Current Limitations}
\label{subsec:current-limitations}

\subsubsection{Absence of Forward Secrecy and Post-Compromise Security}

The most significant cryptographic limitation of the current system is the absence of continuous forward secrecy (FS) and post-compromise security (PCS). Ember uses per-conversation static symmetric keys with explicit rotation, which reduces the key exposure window relative to a purely static design but does not provide the per-message key evolution guarantees offered by Double Ratchet protocols~\cite{tight_security_double_ratchet_2024}. Compromise of the active conversation key exposes all messages encrypted under that key version, both past and future, until rotation is performed. Because rotation is manual and infrequent, the effective compromise window may span a substantial volume of messages.

This limitation is intentional and documented in the threat model (Section~3.4), but it represents the principal gap between Ember's current guarantees and those of mature secure messaging systems such as Signal. It bounds the strength of Ember's confidentiality claims to scenarios in which the conversation key has not been compromised, and it means that Ember cannot claim cryptographic recovery from state exposure without explicit user intervention.

\subsubsection{No Group Messaging or Multi-Device Support}

Ember currently supports only one-to-one conversations. No group messaging semantics, sender-key distribution, or multi-party state management are implemented. Extending the system to support group communication would require addressing scalable key distribution, membership consistency, and epoch-based key evolution---challenges that are well-characterised by the MLS protocol standardisation effort~\cite{rfc9420_mls, authenticated_group_management_mls_2023} but that introduce substantial complexity in serverless environments where globally consistent state and reliable delivery cannot be assumed.

Similarly, Ember does not support multi-device operation. A user's conversation keys and message history are bound to a single device, and there is no mechanism for synchronising cryptographic state across multiple endpoints. Multi-device support in end-to-end encrypted systems typically requires either a device-linking protocol with cross-device key agreement or a centralised device registry, both of which conflict with Ember's serverless design goals.

\subsubsection{Availability Constraints and Offline Delivery}

Ember's serverless architecture imposes a synchronous availability requirement: both peers must be simultaneously reachable on the overlay for message delivery to succeed. There is no store-and-forward capability, no message queueing infrastructure, and no mechanism for deferred delivery to offline recipients. While this constraint is acceptable for a research prototype and is shared by comparable systems such as Ricochet~\cite{ricochet_message_layer_encryption}, it limits Ember's practical utility for asynchronous communication patterns that are routine in everyday messaging.

The foreground service model used for inbound reception mitigates this constraint when the application is running, but Android's increasingly aggressive background execution restrictions may terminate or throttle the listener under certain power-saving states. The dynamic test was conducted under favourable conditions (both devices active, local network), and the system's behaviour under prolonged backgrounding, Doze mode, or OEM-specific process management was not evaluated.

\begin{comment}

\subsubsection{Limited Input Validation and Protocol Hardening}

The OWASP ASVS mapping (Section~\ref{subsec:asvs}) identified input and intent validation as the principal gap area, with weaker coverage in V11 (Data Validation) compared to cryptography and communications controls. Ember's envelope parsing performs basic structural and type checks, but systematic boundary-condition handling, protocol-version negotiation robustness, and defence against malformed or adversarially crafted envelopes have not been subjected to fuzzing or adversarial testing. An attacker capable of delivering crafted envelopes to the listener service could potentially exploit parsing weaknesses to cause denial-of-service or undefined behaviour, even though the HMAC gate would prevent plaintext production from unauthenticated data.

\subsubsection{Absence of Side-Channel and Timing Analysis}

No empirical measurement of side-channel behaviour has been performed. Timing variations in cryptographic operations, memory access patterns during key handling, and observable differences in processing time between valid and invalid messages could theoretically be exploited by a local or co-located adversary. While the use of standard library implementations of AES-GCM, HMAC-SHA256, and HKDF provides some baseline resistance to timing attacks (assuming constant-time implementations in the underlying Android cryptographic provider), this has not been independently verified for Ember's specific execution environment.
\end{comment}
\subsubsection{Platform-Specific Storage Remanence}

As discussed in Section~2.8 and the threat model (Section~3.4), Ember cannot guarantee physical secure deletion of expired message data on Android. TTL-driven cleanup removes records from the application database, but residual data may persist in SQLite write-ahead logs, filesystem journal structures, or untrimmed flash storage~\cite{why_data_deletion_fails_2017, digital_forensics_android_privacy}. This limitation is fundamental to the Android storage stack rather than specific to Ember, but it bounds the strength of Ember's ephemerality claims to application-level access minimisation rather than forensic-grade erasure.

\subsubsection{Test Build Instrumentation as an Endpoint Risk}

The dynamic test reported in Section~\ref{subsec:dynamic-analysis} relied on an instrumented build that emits plaintext message content to the Android logcat stream. While this instrumentation was essential for validating the correlation between user actions and network behaviour, it represents an endpoint-side confidentiality risk if present in any non-test deployment. The current codebase does not enforce build-variant separation between instrumented and release configurations at the compiler level, meaning that suppression of plaintext logging depends on developer discipline rather than structural guarantees. Formalising this separation through build flavour enforcement or compile-time log stripping would reduce the risk of accidental plaintext leakage in production builds.

\subsection{Future Work: Ratcheting-Based Security}
\label{subsec:future-ratcheting}

The most impactful single improvement to Ember's cryptographic posture would be the integration of a ratcheting-based key evolution protocol to provide continuous forward secrecy and post-compromise security. The Double Ratchet, as deployed in Signal and formally analysed by Collins et al.~\cite{tight_security_double_ratchet_2024}, represents the current standard for this class of guarantees. However, incorporating a Double Ratchet into Ember's serverless, peer-to-peer architecture introduces several non-trivial challenges that must be addressed before deployment.

\paragraph{Asynchronous key exchange without pre-key servers.}
The standard Signal deployment model relies on a central server to publish and distribute pre-key bundles, enabling asynchronous key establishment between parties that are not simultaneously online. Ember's serverless architecture eliminates this infrastructure. A viable integration path would need to either (i)~restrict initial key exchange to synchronous sessions where both peers are online, (ii)~introduce a minimal, untrusted pre-key relay analogous to Cwtch's discardable servers~\cite{cwtch_infrastructure}, or (iii)~explore alternative asynchronous key exchange constructions that do not depend on persistent public key directories. Each approach entails distinct trade-offs between availability, metadata exposure, and architectural complexity.

\paragraph{State synchronisation under unreliable delivery.}
Ratcheting protocols require that both parties maintain consistent views of the evolving key state. Message loss, reordering, or duplication---all of which are possible in Ember's per-connection TCP model---can desynchronise ratchet state and cause message decryption failures. Signal addresses this through message counters, skipped message key caches, and server-mediated delivery guarantees. In a peer-to-peer environment without delivery infrastructure, Ember would need to implement robust state recovery mechanisms, potentially including ratchet state checkpointing and resynchronisation protocols, while ensuring that such mechanisms do not themselves introduce new attack surfaces.

\paragraph{Verification complexity.}
Formal analyses of the Double Ratchet~\cite{tight_security_double_ratchet_2024, formal_analysis_signal_loet_2024, formal_analysis_ratchet_mdpi_2025} demonstrate that security guarantees are tightly coupled across protocol stages and that compositional reasoning is difficult. Integrating a ratchet into Ember would therefore require not only correct implementation but also re-evaluation of the threat model, explicit analysis of the composed protocol (ratchet plus Ember's envelope and authentication layers), and ideally some form of formal or mechanised verification. Given the well-documented difficulty of achieving correct ratcheting implementations even in centralised systems, a phased approach---beginning with a symmetric ratchet (chain key evolution) before introducing asymmetric Diffie--Hellman ratchet steps---would reduce implementation risk while providing incremental security improvements.

Ratcheting-based security is therefore treated as the highest-priority future work item, with the understanding that correctness must be established rigorously before any claims of forward secrecy or post-compromise security are made.

\subsection{Future Work: Group Messaging}
\label{subsec:future-group}

Extending Ember to support group conversations would significantly broaden its functional utility but introduces challenges that are qualitatively different from those of two-party communication. The MLS protocol~\cite{rfc9420_mls} provides a well-specified framework for group key management with forward secrecy and post-compromise security, and serves as the primary reference point for any future group messaging extension.

\paragraph{Delivery and consistency semantics.}
MLS assumes the availability of a delivery service that can reliably disseminate messages to all group members and maintain a consistent view of group state~\cite{rfc9420_mls}. In Ember's serverless architecture, no such service exists. A group extension would need to address how messages are distributed to multiple peers (direct fan-out, gossip, or relay-based approaches), how membership changes are propagated consistently, and how the system behaves when some group members are unreachable. Each approach involves trade-offs between latency, bandwidth, consistency, and metadata exposure.

\paragraph{Authentication and identity binding.}
Group messaging requires a mechanism for authenticating group membership and binding long-term identities to cryptographic credentials. MLS delegates this to an external authentication service, which is straightforward in centralised deployments but problematic in a fully decentralised environment. Ember would need to develop a peer-to-peer membership management protocol, potentially leveraging its existing key fingerprint verification mechanism as a building block for group-level trust establishment.

\paragraph{Scalability and state management.}
MLS's tree-based key management structure provides logarithmic scaling for group key updates, but its correctness depends on consistent tree state across all members~\cite{authenticated_group_management_mls_2023}. Maintaining this consistency without a central coordinator is an open research problem, particularly under network partition or churn. For an initial group extension, Ember might adopt a simpler sender-key model (as used by Signal for large groups), accepting weaker forward secrecy properties in exchange for reduced state synchronisation requirements, and migrate to full MLS-style group state as the infrastructure matures.

\subsection{Future Work: Formalisation and Verification}
\label{subsec:future-formal}

The current security assessment of Ember relies on static code analysis, unit testing, OWASP compliance mapping, threat modelling, and dynamic network analysis. While this combination provides multi-source evidence for the system's security properties, it falls short of formal verification. Several directions for formalisation would materially strengthen confidence in Ember's guarantees.

\paragraph{Protocol specification and symbolic analysis.}
Ember's message processing pipeline, key rotation protocol, and envelope authentication scheme could be specified as a formal protocol model and subjected to symbolic analysis using tools such as ProVerif or Tamarin. Such analysis would provide machine-checked guarantees for properties including message confidentiality, authentication, and replay resistance under the Dolev--Yao adversary model. The relatively small protocol surface of Ember (compared to full ratcheting systems) makes it a tractable candidate for this class of analysis.

\paragraph{Fuzzing and adversarial input testing.}
The envelope parser, JSON deserialiser, length-prefix framing handler, and key rotation state machine are all candidates for coverage-guided fuzzing. Fuzzing would complement the existing static analysis by identifying edge cases, boundary-condition failures, and potential denial-of-service vectors that are not detectable through ruleset-driven scanning. Integration of fuzzing into the continuous testing pipeline would provide ongoing regression coverage as the codebase evolves.

\paragraph{Side-channel measurement.}
Empirical measurement of timing behaviour during cryptographic operations, particularly HMAC verification and AES-GCM decryption, would allow assessment of whether Ember is vulnerable to timing-based oracle attacks. While the use of platform-provided cryptographic primitives offers some baseline protection, the interaction between Ember's application-layer processing (HMAC gate, key lookup, error handling) and the underlying cryptographic library has not been profiled for timing leakage.

\paragraph{Instrumented device testing.}
The performance measurements reported in this paper are derived from logcat timestamps and non-instrumented unit tests. Future work should include Android-instrumented benchmarks executed on physical devices under controlled conditions, measuring cryptographic operation latency, database I/O overhead, and end-to-end delivery latency with statistical rigour. Such measurements would provide definitive performance characterisation suitable for comparison with other mobile secure messaging implementations.

\paragraph{Adversarial network testing.}
The dynamic test was conducted under benign network conditions. Future testing should include adversarial scenarios such as packet injection, replay attacks, connection reset injection, malformed envelope delivery, and overlay routing disruption. These tests would validate Ember's defensive processing model under conditions that more closely approximate the active network adversary assumed in the threat model (Section~3.1), and would provide empirical evidence for the robustness of the HMAC gate and envelope validation logic.

\section{Conclusion}
\label{sec:conclusion}

This paper has presented Ember, a serverless peer-to-peer messaging system that provides end-to-end encrypted communication over a decentralised IPv6 mesh network. The system was designed to investigate whether secure messaging can be achieved without centralised infrastructure, under conditions where regulatory and architectural pressures increasingly threaten the viability of conventional end-to-end encryption deployments. Rather than proposing a novel cryptographic primitive, the work focused on system-level composition: integrating established cryptographic mechanisms into a coherent, analysable, and practically deployable architecture on mobile platforms.

Ember makes five concrete, implemented contributions. First, it demonstrates that serverless, peer-to-peer messaging over an encrypted overlay network is technically viable on Android, with message delivery latencies in the range of 77--414~ms under steady-state conditions and cryptographic processing overhead of under 20~ms per message. Second, it enforces a ciphertext-only persistence model in which plaintext message content is never written to disk, materially reducing the forensic value of stored application state. Third, it implements TTL-driven message expiration through WorkManager-based cleanup, providing application-level ephemerality that limits long-term data retention without overstating guarantees about physical storage erasure. Fourth, it provides an explicit key rotation protocol with HKDF-based derivation and mutual peer confirmation, enabling controlled key evolution without server coordination. Fifth, it adopts a layered architecture with explicit trust boundaries that separates UI logic, cryptographic operations, persistence, and network transport, making security-critical paths independently reviewable and testable.

The security assessment combined five complementary evaluation methods: static code analysis (zero high/critical findings across 66 files), unit-test validation of the full cryptographic pipeline (7/7 tests passed), OWASP ASVS v4.0 compliance mapping (77\% alignment at L2 with partial L3 coverage), structured threat modelling across eight adversary classes, and dynamic network security analysis using on-the-wire packet capture correlated with device-level runtime telemetry. The dynamic test provided the strongest empirical evidence for Ember's core confidentiality claims: exhaustive byte-level search of the packet capture for all known plaintext strings returned zero matches, HMAC verification succeeded for all eight messages exchanged between two physical devices, and the verify-then-decrypt processing order was confirmed for every inbound message without exception. No plaintext, JSON envelope structures, or readable application data were recoverable from the captured network traffic.

The limitations of the current system are explicitly acknowledged and precisely bounded. Ember does not provide continuous forward secrecy or post-compromise security; its per-conversation symmetric keys with explicit rotation reduce but do not eliminate the key exposure window. It does not support group messaging, multi-device operation, or asynchronous offline delivery. It does not claim metadata anonymity against traffic analysis adversaries, nor does it guarantee physical secure deletion on Android storage media. Input validation coverage, side-channel resilience, and adversarial protocol robustness have not been empirically tested. These limitations are documented in the threat model and carried transparently through the evaluation, ensuring that Ember's claims are commensurate with its evidence.

Despite these limitations, Ember occupies a defensible position in the secure messaging design space. It demonstrates that the elimination of centralised infrastructure does not preclude strong content confidentiality, authenticated message integrity, or aggressive data minimisation. By avoiding third-party push notification services, central metadata aggregation points, and server-mediated key management, Ember removes well-documented classes of privacy leakage that persist in mainstream encrypted messaging systems. The trade-offs it accepts---synchronous availability, manual key rotation, foreground service resource costs---are explicit, bounded, and appropriate for a research prototype that prioritises correctness and architectural clarity over feature completeness.

The most impactful directions for future work are the integration of ratcheting-based key evolution to provide forward secrecy and post-compromise security, formal protocol verification using symbolic analysis tools, and adversarial testing of the envelope parser and transport layer under hostile network conditions. Each of these directions builds directly on the architectural foundation established by the current system and would extend Ember's guarantees into domains that are currently treated as explicit non-goals.

Ember is offered not as a replacement for mature, widely deployed secure messaging platforms, but as a rigorous exploration of an alternative point in the design space---one in which no central authority need be trusted, no persistent infrastructure need be maintained, and no third-party service need be consulted for communication to occur. Within the boundaries of its stated threat model, the system achieves what it set out to achieve: serverless, peer-to-peer, end-to-end encrypted messaging with minimal retained state, empirically validated confidentiality, and security properties that are explicit, bounded, and defensible.

\acknowledgement{Not applicable}

\funding{The author(s) received no specific funding for this study.}

\authorcontributions{
The authors confirm contribution to the paper as follows: Conceptualization, Hamish Alsop and Leandros Maglaras; methodology, Hamish Alsop; software, Hamish Alsop; validation, Hamish Alsop and Naghmeh Moradopoor; formal analysis, Hamish Alsop; investigation, Hamish Alsop and Leandros Maglaras; resources, Hamish Alsop; data curation, Hamish Alsop and Naghmeh Moradpoor; writing---original draft preparation, Hamish Alsop; writing---review and editing, Hamish Alsop, Leandros Maglaras and Naghmeh Moradpoor; visualization, Hamish Alsop; supervision, Leandros Maglaras  project administration, Hamish Alsop; funding acquisition, Leandros Maglaras. All authors reviewed and approved the final version of the manuscript.}

\availabilityofdataandmaterials{The data that support the findings of this study are available from the Corresponding Author, H.A., upon reasonable request.}

\ethicsapproval{Not applicable.}

\conflictsofinterest{The authors declare no conflicts of interest} 

\bibliography{TSP_template}

\end{document}

%% file: Definitions/package.tex
\usepackage[normalem]{ulem} % For using the command \uuline{} in the \abstract 
\usepackage{amsfonts} % Use for some special math mark
\usepackage{mathcomp} % For permille mark
\usepackage{CJKutf8} % For Chinese font
\usepackage{pifont} % For some special mark 
\usepackage{bm} % Forr math enviroment bold format
\usepackage{bbm} % For some math fancy characters
\graphicspath{{./Definitions/}} % Use for import path the figures paper used

\makeatletter
\def\T@n@@nc@d@ngM@cr@M@d{}
\def\LY@n@@nc@d@ngM@cr@M@d{}
\makeatother

\let\orignewcommand\newcommand  % Store the original \newcommand
\let\newcommand\providecommand  % Make \newcommand behave like \providecommand
\usepackage{verse}
\let\newcommand\orignewcommand  % Use the original `\newcommand` in future

\newsavebox\foobox

\renewcommand{\dagger}{\mathchar"2279}
\renewcommand{\ddagger}{\mathchar"227A}

\newcommand{\mmathit}[1]{
  \ifthenelse{\equal{#1}{\ln}}{\mathit{ln}}{
    \ifthenelse{\equal{#1}{\max}}{\mathit{max}}{\mathit{#1}}
  }
}
\makeatother
\robustify{\footnote}

%% file: TSP_template.bib
@inproceedings{sok_secure_messaging_2015,
  title     = {{SoK}: Secure Messaging},
  author    = {Unger, Nik and Dechand, Sergej and Bonneau, Joseph and Fahl, Sascha and Perl, Henning and Goldberg, Ian and Smith, Matthew},
  booktitle = {2015 IEEE Symposium on Security and Privacy},
  pages     = {232--249},
  year      = {2015},
  publisher = {IEEE}
}

@article{akalin2026chat,
  title   = {From Chat Control to Robot Control: Implications of the Chat Control Proposal for Human-Robot Interaction},
  author  = {Akalin, Neziha and Giaretta, Alberto},
  journal = {arXiv preprint arXiv:2601.02205},
  year    = {2026}
}

@article{abelson2024bugs,
  title     = {Bugs in our pockets: the risks of client-side scanning},
  author    = {Abelson, Harold and Anderson, Ross and Bellovin, Steven M and Benaloh, Josh and Blaze, Matt and Callas, Jon and Diffie, Whitfield and Landau, Susan and Neumann, Peter G and Rivest, Ronald L and Schiller, Jeffrey I and Schneier, Bruce and Teague, Vanessa and Troncoso, Carmela},
  journal   = {Journal of Cybersecurity},
  volume    = {10},
  number    = {1},
  pages     = {tyad020},
  year      = {2024},
  publisher = {Oxford University Press}
}

@article{alsop2025innovating,
  title   = {Innovating Augmented Reality Security: Recent {E2E} Encryption Approaches},
  author  = {Alsop, Hamish and Maglaras, Leandros and Janicke, Helge and Sarker, Iqbal H and Ferrag, Mohamed Amine},
  journal = {arXiv preprint arXiv:2509.10313},
  year    = {2025}
}

@article{slr_secure_instant_messaging_2024,
  title     = {A systematic literature review of secure instant messaging applications from a digital forensics perspective},
  author    = {Onik, Abdur Rahman and Brown, Joseph and Walker, Clinton and Baggili, Ibrahim},
  journal   = {ACM Computing Surveys},
  volume    = {57},
  number    = {9},
  pages     = {1--36},
  year      = {2025},
  publisher = {ACM New York, NY}
}

@inproceedings{tight_security_double_ratchet_2024,
  title     = {On the tight security of the double ratchet},
  author    = {Collins, Daniel and Riepel, Doreen and Tran, Si An Oliver},
  booktitle = {Proceedings of the 2024 ACM SIGSAC Conference on Computer and Communications Security},
  pages     = {4747--4761},
  year      = {2024}
}

@article{formal_analysis_signal_loet_2024,
  title     = {Formal analysis of {Signal} protocol based on logic of events theory},
  author    = {Li, Zehuan and Xiao, Meihua and Xu, Ruihan},
  journal   = {Scientific Reports},
  volume    = {14},
  number    = {1},
  pages     = {20606},
  year      = {2024},
  publisher = {Nature Publishing Group UK London}
}

@article{formal_analysis_ratchet_mdpi_2025,
  title     = {Formal Analysis of Ratchet Protocols Based on Logic of Events},
  author    = {Xiao, Meihua and Wan, Hongbin and Fan, Hongming and Shao, Huaibin and Li, Zehuan and Yang, Ke},
  journal   = {Applied Sciences},
  volume    = {15},
  number    = {13},
  pages     = {6964},
  year      = {2025},
  publisher = {MDPI}
}

@article{x3dh_efficient_construction_2022,
  title     = {An efficient and generic construction for {Signal's} handshake ({X3DH}): post-quantum, state leakage secure, and deniable},
  author    = {Hashimoto, Keitaro and Katsumata, Shuichi and Kwiatkowski, Kris and Prest, Thomas},
  journal   = {Journal of Cryptology},
  volume    = {35},
  number    = {3},
  pages     = {17},
  year      = {2022},
  publisher = {Springer}
}

@article{strongly_deniable_ake_2016,
  title   = {Improved strongly deniable authenticated key exchanges for secure messaging},
  author  = {Unger, Nik and Goldberg, Ian},
  journal = {Proceedings on Privacy Enhancing Technologies},
  volume  = {2018},
  number  = {1},
  pages   = {21--66},
  year    = {2018}
}

@misc{rfc9420_mls,
  title  = {The {Messaging Layer Security (MLS)} Protocol},
  author = {Barnes, Richard and Beurdouche, Benjamin and Robert, Raphael and Millican, Jon and Omara, Emad and Cohn-Gordon, Katriel},
  year   = {2023},
  note   = {RFC 9420, IETF}
}

@inproceedings{authenticated_group_management_mls_2023,
  title     = {{TreeSync}: Authenticated Group Management for {Messaging Layer Security}},
  author    = {Wallez, Th{\'e}ophile and Protzenko, Jonathan and Beurdouche, Benjamin and Bhargavan, Karthikeyan},
  booktitle = {32nd USENIX Security Symposium (USENIX Security 23)},
  pages     = {1217--1233},
  year      = {2023}
}

@inproceedings{metadata_privacy_beyond_tunneling_2024,
  title     = {Metadata Privacy Beyond Tunneling for Instant Messaging},
  author    = {Nelson, Boel and Pagnin, Elena and Askarov, Aslan},
  booktitle = {2024 IEEE 9th European Symposium on Security and Privacy (EuroS\&P)},
  pages     = {697--723},
  year      = {2024},
  publisher = {IEEE}
}

@misc{cwtch_infrastructure,
  title  = {Cwtch: Privacy Preserving Infrastructure for Asynchronous, Decentralized, Multi-Party and Metadata Resistant Applications},
  author = {Lewis, Sarah Jamie},
  year   = {2018},
  note   = {Open Privacy Research Society}
}

@misc{ricochet_message_layer_encryption,
  title  = {Message-Layer Encryption in {Ricochet}},
  author = {Kirsh, Liam B},
  year   = {2017}
}

@article{survey_mesh_messaging_2021,
  title   = {Survey of mesh networking messengers},
  author  = {Bl{\"o}chinger, Simon and von Seck, Richard},
  journal = {Network},
  volume  = {1},
  pages   = {163--181},
  year    = {2021},
  publisher = {MDPI}
}

@misc{cryptography_in_the_wild_briar,
  title  = {Cryptography in the Wild: An Analysis of {Briar}},
  author = {Song, Yuanming},
  school = {ETH Zurich},
  year   = {2022}
}

@misc{matrix_yggdrasil_experiment,
  title  = {Experimenting with {Matrix} Federation over {Yggdrasil}},
  author = {Floury, Timoth{\'e}e},
  school = {EPFL},
  year   = {2019}
}

@inproceedings{yggdrasil_routing_scheme,
  title     = {Yggdrasil Routing Scheme as a Basis for Large-Scale Decentralized Networking},
  author    = {Pestov, Oleksiy and Kyrychek, Heorhii},
  booktitle = {CEUR Workshop Proceedings},
  volume    = {3790},
  year      = {2024},
  note      = {ICST-2024}
}

@inproceedings{android_data_residue_attacks_2017,
  title     = {All Your Droid Are Belong To Us: A Survey of Current {Android} Attacks},
  author    = {Vidas, Timothy and Votipka, Daniel and Christin, Nicolas},
  booktitle = {WOOT},
  year      = {2011}
}

@article{why_data_deletion_fails_2017,
  title     = {Secure Deletion on Log-structured File Systems},
  author    = {Reardon, Joel and Capkun, Srdjan and Basin, David},
  journal   = {ACM Transactions on Storage},
  volume    = {12},
  number    = {2},
  pages     = {1--32},
  year      = {2016},
  publisher = {ACM}
}

@article{digital_forensics_android_privacy,
  title     = {Forensic analysis of {WhatsApp Messenger} on {Android} smartphones},
  author    = {Anglano, Cosimo},
  journal   = {Digital Investigation},
  volume    = {11},
  number    = {3},
  pages     = {201--213},
  year      = {2014},
  publisher = {Elsevier}
}

@article{forensic_ephemeral_messaging,
  title     = {Forensic analysis of ephemeral messaging applications: Disappearing messages or evidential data?},
  author    = {Heath, Howard and MacDermott, {\'A}ine and Akinbi, Alex},
  journal   = {Forensic Science International: Digital Investigation},
  volume    = {46},
  pages     = {301585},
  year      = {2023},
  publisher = {Elsevier}
}

@article{user_perceptions_deletion,
  title     = {Exploring User Perceptions of Deletion in Mobile Instant Messaging Applications},
  author    = {Schnitzler, Theodor and Utz, Christine and Farke, Florian M. and P{\"o}pper, Christina and D{\"u}rmuth, Markus},
  journal   = {Journal of Cybersecurity},
  volume    = {6},
  number    = {1},
  pages     = {tyz016},
  year      = {2020},
  publisher = {Oxford University Press}
}

@article{push_notification_leakage,
  title   = {The Medium is the Message: How Secure Messaging Apps Leak Sensitive Data to Push Notification Services},
  author  = {Samarin, Nikita and Sanchez, Alex and Chung, Trinity and Dan, Akshay and Juleemun, Bhavish and Gilsenan, Conor and Merrill, Nick and Reardon, Joel and Egelman, Serge},
  journal = {Proceedings on Privacy Enhancing Technologies},
  volume  = {2024},
  number  = {4},
  pages   = {967--982},
  year    = {2024}
}

@article{android_key_storage_analysis_2025,
  title   = {{KeyDroid}: A Large-Scale Analysis of Secure Key Storage in {Android} Apps},
  author  = {Blessing, Jenny and Anderson, Ross J. and Beresford, Alastair R.},
  journal = {arXiv preprint arXiv:2507.07927},
  year    = {2025}
}
